\documentclass[aps,prb,amsmath,amssymb,twocolumn,superscriptaddress,showpacs,floatfix]{revtex4-1}

\usepackage{ifpdf}
 \newif\ifpdf
\ifx\pdfoutput\undefined
   \pdffalse
\else
   \pdfoutput=1
   \pdftrue
\fi
\ifpdf
   \usepackage{graphicx}
   \usepackage{epstopdf}
\else
   \usepackage{graphicx}
\fi

\DeclareMathOperator{\TCFE}{\mathit{T}_{\mathrm{\scriptscriptstyle C}}^{\mathrm{\scriptscriptstyle FE}}}

\DeclareMathOperator{\TCFM}{\mathit T_{\mathrm{\scriptscriptstyle C}}^{\mathrm{\scriptscriptstyle FM}}}

\usepackage{graphicx}
\usepackage{dcolumn}
\usepackage{bm}

\usepackage{amsmath}
\usepackage{amssymb}

\begin{document}


\title{Phenomenological theory of magneto-electric coupling in granular multiferroics}

\author{O.~G.~Udalov}
\affiliation{Department of Physics and Astronomy, California State University Northridge, Northridge, CA 91330, USA}
\affiliation{Institute for Physics of Microstructures, Russian Academy of Science, Nizhny Novgorod, 603950, Russia}

\author{N.~M.~Chtchelkatchev}
\affiliation{Department of Physics and Astronomy, California State University Northridge, Northridge, CA 91330, USA}
\affiliation{L.D. Landau Institute for Theoretical Physics, Russian Academy of Sciences,117940 Moscow, Russia}
\affiliation{Department of Theoretical Physics, Moscow Institute of Physics and Technology, 141700 Moscow, Russia}

\author{I.~S.~Beloborodov}
\affiliation{Department of Physics and Astronomy, California State University Northridge, Northridge, CA 91330, USA}

\date{\today}


\begin{abstract}
We study coupling between the ferroelectric polarization and magnetization of granular
ferromagnetic film using a phenomenological model of combined multiferroic system consisting of granular ferromagnetic
film placed above the ferroelectric (FE) layer. The coupling is due to screening of Coulomb interaction in the
granular film by the FE layer. Below the FE Curie temperature the magnetization has
hysteresis as a function of electric field. Below the magnetic ordering temperature the polarization
has hysteresis as a function of magnetic field. We study the magneto-electric coupling
for weak and strong spatial dispersion of the FE layer. The effect of mutual influence
decreases with increasing the spatial dispersion of the FE layer.
For weak dispersion the strongest coupling occurs in the vicinity of the ferroelectric-paraelectric phase transition.
For strong dispersion the situation is the opposite.
We study the magneto-electric coupling as a function of distance between the FE layer and
the granular film. For large distances the coupling decays exponentially due to the
exponential decrease of electric field produced by the oscillating charges in the granular ferromagnetic film.

\end{abstract}

\maketitle

\section{Introduction\label{sec:intro}}

Currently the field of multiferroics is a very active area of research.~\cite{Scott2006, Spal2007, Bar2008, Ohtani2000, Ohno2008, Ohno2003} It promises numerous applications, but provides much more fundamental challenges. Vast variety of different multiferroic materials are currently studied
by many groups who are looking for strong magneto-electric (ME) coupling. Among them are single crystals possessing intrinsic ME coupling,~\cite{Bal2005, Ser2006} and composite multiferroics consisting of ferroelectric (FE) and ferromagnetic (FM) layers coupled due to strain or surface charges.~\cite{Nan1994, Schultz2007, Goennenwein2010, Givord2007, Oda2009, Tsymbal2008, Tsymbal2010, Barthelemy2010, Berakdar2009, Berakdar2011}

Recently, granular multiferroics - materials consisting of
magnetic particles embedded into FE matrix attract much attention. A novel mechanism of ME coupling
involving the interplay of the Coulomb blockade effects, intergrain exchange interaction and ferroelectric dielectric response was proposed for these materials~\cite{Beloborodov2014_ME, Beloborodov2014_ME1}. This mechanism was studied using the microscopic theory. In particular, it was shown
that the magnetization of granular multiferroics strongly depends on the FE state leading to the
appearance of an additional magnetic phase transition in the vicinity of the
FE Curie point and to the possibility of controlling the magnetic state of the system by an electric field.
\begin{figure}
\includegraphics[width=0.9\columnwidth]{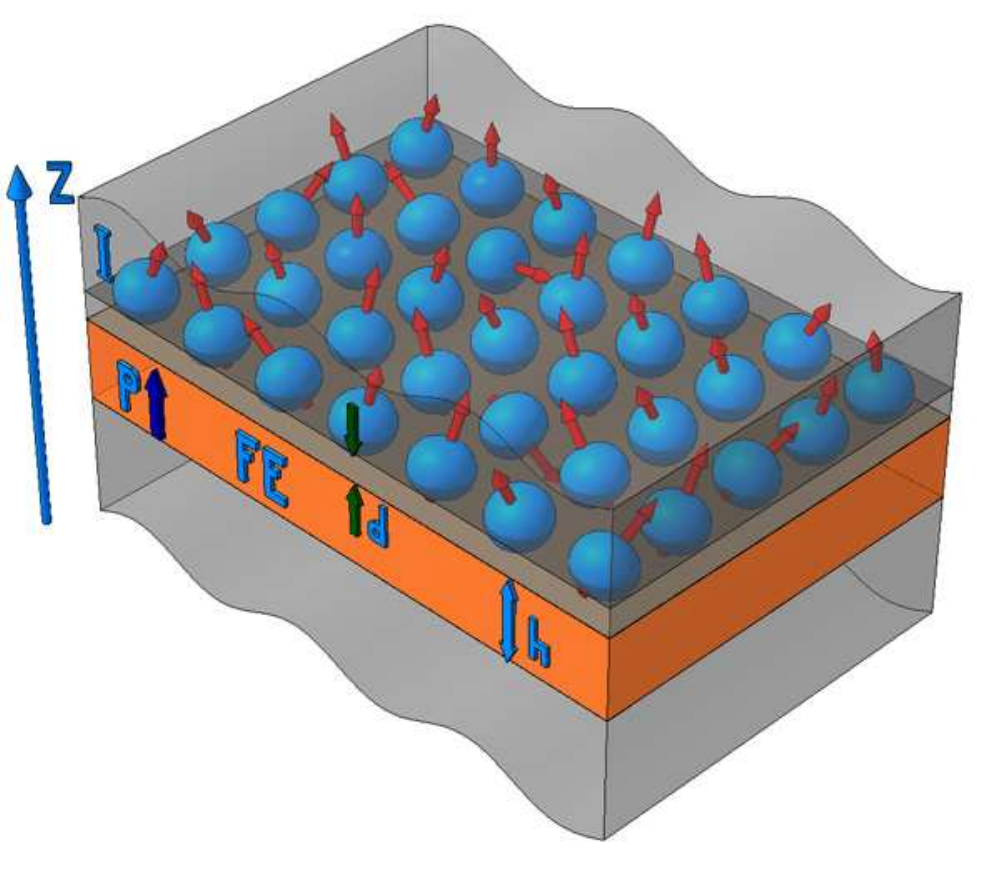}
\caption{(Color online) Composite multiferroic - material consisting of granular ferromagnetic film
placed at distance $d$ above the FE layer (FE) of thickness $h$. Ferromagnetic film consists of
ferromagnetic metallic particles (blue spheres) with finite magnetic moments (red arrows)
embedded into an insulating matrix (I). FE layer has a polarization $P$ along the z-axis.}\label{Fig_Sys}
\end{figure}

In this paper we study the ME coupling mechanism in combined granular multiferroic - material consisting of granular ferromagnetic film (GFM) placed above the FE layer at distance $d$, see Fig.~\ref{Fig_Sys}. In contrast
to the previous works, here we use a phenomenological approach. This approach allows to account
for i) the spatial dispersion of the FE layer and ii) the influence of magnetic subsystem on the FE polarization.
Both these effects were not discussed before.

According to Ref.~\onlinecite{Beloborodov2014_ME} the coupling between the GFM film and the
FE layer occurs due to screening of Coulomb interaction in the GFM film
by the FE layer. The screening was discussed assuming that the
FE layer is a dielectric with a local response. In this case the ME effect has a peak in the vicinity of the
FE Curie temperature. However, real FEs have domain walls of finite thickness increasing
with approaching the paraelectric-ferroelectric phase transition.
The FE layer can not effectively screen the electric field with characteristic
spatial length being smaller than the thickness of the FE domain wall.
The characteristic scale for the electric field produced by the GFM film is defined by the intergrain distance. For
FE domain wall thickness exceeding this scale the coupling between the FE and the GFM layers
is suppressed. This leads to the decrease of the ME effect in the vicinity of
the paraelectric-ferroelectric phase transition, contrary to the local response case.
Such a behavior was not discussed before since the ME effect in the GMF film was studied assuming the
local response of the FE layer. In this paper we study the
influence of the FE spatial dispersion on coupling between the FE layer and the GFM film.

We use a phenomenological approach to study the system with spatial dispersion.
Usually the ME coupling effects are treated using terms proportional to the product of
polarization and magnetization, $\sim \alpha_{\mathrm{me}}P^nM^n$.~\cite{Chupis1982} We describe our
system using three phenomenological parameters: 1) the FE polarization, 2) the GFM film
magnetization, and 3) the spatial oscillations of charge in the GFM. The later parameter is crucial for granular materials
since these materials have complicated morphology leading to the
inevitable formation of charge oscillations. We use the local quadrupole moment to describe
the system since the average polarization and the average charge in the granular film is zero.
The microscopic theory of ME coupling in GMF shows that
the charge oscillations are responsible for this coupling, thus supporting
the use of these three parameters.

Phenomenologically the influence of the FE subsystem on the magnetic subsystem
is described by the term involving both polarization and magnetization
in the total energy of the system.~\cite{Chupis1982} This contribution
leads to the inverse effect - the influence of magnetic subsystem on the FE subsystem.
This effect will be discussed in the present paper.

The paper is organized as follows. In Sec.~\ref{sec:Model} we discuss
the model for combined granular multiferroic system. Using this model we
consider two cases of weak and strong spatial dispersion of the FE layer in Secs.~\ref{Sec:WD} and \ref{Sec:SD}, respectively.
In Sec.~\ref{Sec:Micro} we discuss the phenomenological and microscopic approaches.
Finally, we consider the validity of our approach in Sec.~\ref{Sec:Disc}

\section{The model}\label{sec:Model}

\subsection{System parameters}

In this section we discuss the model of composite multiferroics - materials consisting of
two thin layers: i) ferroelectric (FE) layer and ii) granular ferromagnetic film (GFM) made of ferromagnetic grains embedded into an insulating matrix, see Fig.~\ref{Fig_Sys}. The grains have average radius $a$ of few nm with the distance between
grains being $1-2$ nm. The distance between the neighbouring grain centres is $L_\mathrm g$. Each grain is
characterized by large Curie temperature, much larger than all other characteristic energy scales in the problem.
Therefore, each particle is in the FM state. Due to the interaction between particles the macroscopic
FM state may occur in the GFM film for temperatures $T < \TCFM$, where $\TCFM$ is the ferromagnetic ordering temperature.
\begin{figure}
\includegraphics[width=0.7\columnwidth]{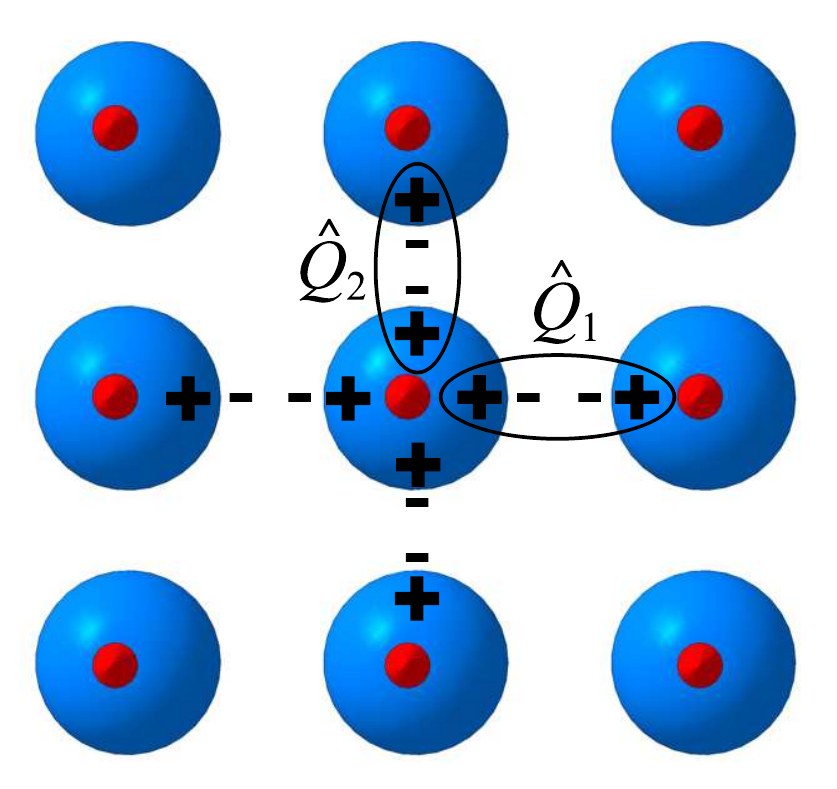}\\
(a)
\includegraphics[width=1\columnwidth]{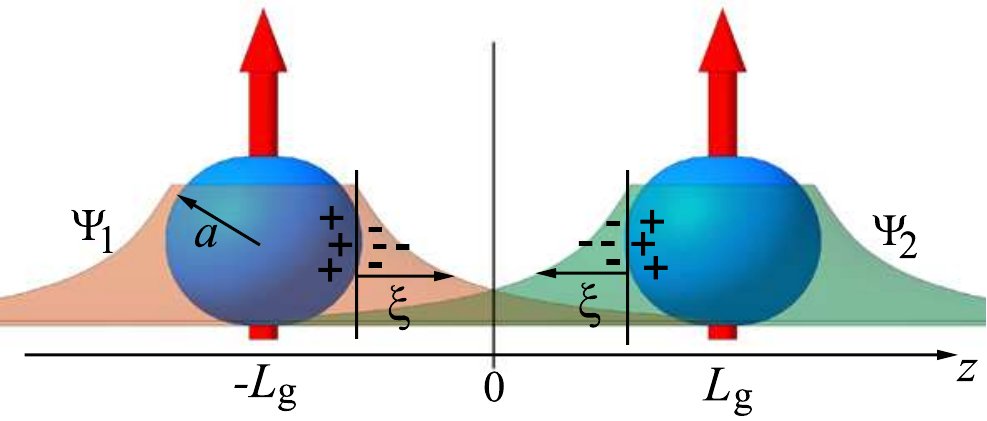}
(b)
\caption{(Color online) (a) Lattice of metallic grains.
The interparticle spacing has a small negative and grains have small positive charges
due to electron tunneling between grains leading to the formation of quadrupole moments in the regions between grains.
There are two types of quadrupoles, $\hat Q_1$ and $\hat Q_2$. (b) Two magnetic grains. Electron wave functions ($\Psi_1$ and $\Psi_2$) extend
beyond the grains and overlap in the region between the grains. $\xi$ is the decay length of
the electron wave functions. Quadrupole moment appears due to presence of electrons outside the grains.
Exchange interaction between grains appears due to the overlap of electron wave functions.}\label{Fig_Quadr}
\end{figure}

There are three phenomenological parameters characterizing the system:
1) the coordinate dependent electric polarization of the FE layer, $\mathbf{P}$;
2) the average magnetization of the GFM layer, $\mathbf{M}$;
3) the spatial oscillations of electric charge in the GFM film appearing due to inhomogeneous
distribution of metallic inclusions in the granular film, see Fig.~\ref{Fig_Quadr}. Even for equal
number of electrons and ions in a certain grain, the electron wave functions extend beyond the
metallic grains leading to the appearance of a non-zero local electric dipole moment.
Opposite dipole moments of two neighbouring grains form a quadrupole moment
between each pair of grains. Therefore the system is described by the ensemble of quadrupoles
with moments $\hat Q_i$, see Fig.~\ref{Fig_Quadr}.

In addition, the system is characterized by several length scales.
The domain wall thickness $L_{\mathrm p}$ in the FE away from the
transition point can be comparable with interatomic distance. In this case $L_{\mathrm p}$
is smaller than the intergrain distance $L_\mathrm g$. Close to the transition point the situation
is the opposite, $L_{\mathrm p} > L_\mathrm g$.
The magnetic domain wall thickness $L_{\mathrm m}$ in the GFM film is much
larger than the intergrain distance, $L_{\mathrm m} >  L_\mathrm g$.

\subsection{Free energy}

The total free energy of the system consists of three contributions: 1) the energy of the FE layer, $W^{\mathrm{FE}}$,
2) the energy of the GFM film, $W^{\mathrm{GFM}}$, and 3) the interaction energy between two subsystems, $W^{\mathrm I}$.
Below we discuss each energy contribution in details.

\subsubsection{Energy of granular ferromagnetic film, $W^{\mathrm{GFM}}$}

The free energy of GFM film, $W^{\mathrm{GFM}}$ has two contributions
\begin{equation}\label{VolEnM_1}
W^{\mathrm{GFM}}=W^{\mathrm{GFM}}_{\mathrm{m}}+W^{\mathrm{GFM}}_{\mathrm{c}},
\end{equation}
where $W^{\mathrm{GFM}}_{\mathrm{m}}$ is the energy of magnetic subsystem~\cite{landauVol8}
\begin{equation}\label{VolEnM_2}
W^{\mathrm{GFM}}_{\mathrm{m}}=\alpha_{\mathrm{M}}M^2+
\beta_{\mathrm{M}}M^4-(\mathbf{M}\cdot\mathbf{B})+\delta_{\mathbf{M}}(\nabla\mathbf{M})^2. \end{equation}
Here $\alpha_{\mathrm{M}}$, $\beta_{\mathrm{M}}$, $\delta_{\mathrm{M}}$ are some phenomenological constants and
$\mathbf{B}$ is the external magnetic field.

The second contribution, $W^{\mathrm{GFM}}_{\mathrm{c}}$ in Eq.~(\ref{VolEnM_1}) is due to spatial charge oscillations.
The simplest model of regular rectangular array of identical grains with the lattice parameter $L_\mathrm g$ is characterized by the regular array of quadrupoles $\hat Q^i$ which can be characterized by magnitude $Q^i=Q^i_{xx}+Q^i_{yy}$. Below we consider
a uniform spatial distribution of quadrupole moments and
introduce a single parameter describing the system of quadrupoles, $Q$ ($Q^i=Q$). There are two types
of quadrupoles, $\hat Q^1$ and $\hat Q^2$, see Fig.~\ref{Fig_Quadr}. These quadrupoles are  transformable
one into another using the rotation $\pi/2$ ($Q_{xx}^1=Q^2_{yy}$, $Q_{yy}^1=Q^2_{xx}$).
Both quadrupoles have the same magnitude $Q$, however the electric field produced by these
quadrupoles is different.

We define the electrical induction of electric field produced by quadrupole with unit moment ($Q=1$)
as $\mathbf D^{\mathrm q}_i(\mathbf r, \mathbf r_i)$, where index $i$ stands for quadrupole $i$,
$\mathbf r_i$ denotes the quadrupole position, and $\mathbf r$ defines the observer position.
Below we will omit vectors $\mathbf r_i$ for simplicity keeping the index $i$ only.
There are two different spatial distributions of electric field $\mathbf D^{\mathrm q}_i$
corresponding to two types of quadrupoles. The total electric field
produced by quadrupoles is $\mathbf D=Q\sum_i \mathbf D^{\mathrm q}_i$.

The phenomenological parameter $\hat Q_i$ is different from
the polarization $\mathbf P$ and magnetization $\mathbf M$ since quadrupoles
appear due to complex morphology and not due to a phase transition.
In the absence of magnetization and ferroelectricity the quadrupoles are
described by the following free energy  $W^{0}_{\mathrm{c}}=\alpha_{Q}(Q-Q_0)^2$,
where $Q_0$ is the equilibrium magnitude of quadrupoles at a given temperature $T$ and
parameter $\alpha_{Q}$ depends on temperature.

Quadrupoles interact with each other via electric field. The energy density of this field is
\begin{equation}\label{Eq_Quad_EnQQ}
W^{\mathrm E}=\frac{Q^2}{8\pi\Omega_{\mathrm{GFM}}}\sum_{i,j}\int{d^3r \mathbf D^{\mathrm q}_i(\mathbf r)\mathbf D^{\mathrm q}_j(\mathbf r)},
\end{equation}
where $\Omega_{\mathrm{GFM}}$ is the volume of the GFM film. Without loss of generality we assume that beside the FE
layer dielectric permittivity of all over space is approximately 1.
The average electric field produced by the ensemble of quadrupoles is zero.
Therefore the interference of external field $\mathbf E_0$ and the quadrupole field
is negligible, $\int{d^3r\mathbf E_0\cdot\mathbf \sum_i D^{\mathrm q}_i}=0$.

The spatial charge oscillations produce an additional contribution to the system Coulomb energy $W^{\mathrm{GFM}}$.
This contribution defines the coupling
between quadrupoles and magnetic subsystem. The exchange interaction is the
short range interaction. Thus we use the local coupling between parameter $Q$ and magnetization $M$.
Since $Q$ is invariant with respect to the spatial inversion it enters linearly into the coupling term.
Finally, we obtain the following result for the energy of quadrupoles
\begin{equation}\label{Eq_Quad_EnQM}
W^{\mathrm{GFM}}_{\mathrm{c}}=W^{0}_{\mathrm{c}}+W^{\mathrm E}+\gamma(Q-Q_0)M^2,
\end{equation}
where $\gamma$ is a phenomenological parameter. The higher order terms, $\sigma^4M^2$, $\sigma^2M^4$, and $\sigma^4M^4$
can be taken into account as well. For simplicity we consider only the lowest order coupling term
between $Q$ and $M$. The microscopic origin of this coupling is discussed in Sec.~\ref{Sec:Micro}.

\subsubsection{Energy of ferroelectric layer, $W^{\mathrm{FE}}$}

The free energy of the FE layer has the form,~\cite{devonshire1949xcvi, Levan1983, Tilley2001, chandra2007landau}
\begin{equation}\label{Eq_En_FE}
W^{\mathrm{FE}}=\alpha_{\mathrm{P}}P^2+\beta_{\mathrm{P}}P^4+\delta_{\mathbf{P}}(\nabla\mathbf{P})^2 -(\mathbf{P}\cdot\mathbf{E}_0).
\end{equation}
Here $\alpha_{\mathrm{P}}$, $\delta_{\mathbf{P}}$ and $\beta_{\mathrm{P}}$ are phenomenological constants and
$\mathbf{E}_0$ is the homogeneous external electric field directed perpendicular to the FE layer (z-axis).

We notice that the charges responsible for the
external field $\mathbf E_0$ and quadrupoles in the GFM film have a different origin:
the charges outside the GFM film are created by the voltage source
leading to the fixed electric field $\mathbf E_0$ but not to the
fixed electric induction $\mathbf D_0$ while the quadrupoles appear
due to complex morphology producing a finite electric
field induction $\mathbf D$ rather than the electric field $\mathbf E$.

\subsubsection{Interaction energy between two subsystems, $W^{\mathrm I}$}

The coupling between the FE layer and
the GFM film occurs due to the interaction of electric field produced by quadrupoles
in the GFM film with the FE layer
\begin{equation}\label{Eq_Quad_EnQP}
W^{\mathrm I}=-\frac{Q}{2\Omega_{\mathrm{GFM}}}\sum_{i}\int{d^3r \mathbf D^{\mathrm q}_i(\mathbf r)\mathbf P(\mathbf r)},
\end{equation}
where the FE polarization has the form
\begin{equation}\label{Eq_Pol_Gen}
\mathbf P=\mathbf P_0+\mathbf P^{(1)}(\mathbf r)+\mathbf P^{(2)}(\mathbf r).
\end{equation}
Here $\mathbf P_0$ is the spontaneous (or external field induced) uniform polarization of the
FE layer. It depends on the external field below and above the transition temperature $\TCFE$.
We assume that the electric field created by quadrupoles in the FE layer is weak.
The terms $\mathbf P^{(1,2)}(\mathbf r)$ in Eq.~(\ref{Eq_Pol_Gen}) are the linear and quadratic responses of the
FE to the quadrupoles field $\mathbf D$
\begin{equation}\label{Eq_Pol_GenL}
\mathbf P^{(1)}(\mathbf r)=Q\sum_i\int_{\Omega_{\mathrm{FE}}}{d^3r' \hat\chi(\mathbf r, \mathbf r')\mathbf D^{\mathrm q}_i(\mathbf r')},
\end{equation}
where $\hat\chi(\mathbf r, \mathbf r')$ is the linear response function of the FE layer
to the electric induction. In general, $\hat\chi(\mathbf r, \mathbf r')$ is a tensor depending
on the polarization $\mathbf P_0$, temperature, and external electric field $\mathbf E_0$. Inside
the FE layer $\hat\chi(\mathbf r, \mathbf r')$ depends on both coordinates
$\mathbf r$ and $\mathbf r'$ due to boundary conditions. In the bulk the susceptibility
depends only on the coordinate difference $(\mathbf r- \mathbf r')$.

The quadratic response in Eq.~(\ref{Eq_Pol_Gen}) has the form
\begin{equation}\label{Eq_Pol_GenQ}
\mathbf P^{(2)}(\mathbf r) =Q^2\sum_{i,j}\int_{\Omega_{\mathrm{FE}}}{d^3r'd^3r'' \hat\chi^{(2)}(\mathbf r, \mathbf r',\mathbf r'')\mathbf D^{\mathrm q}_i(\mathbf r')\mathbf D^{\mathrm q}_j(\mathbf r'')},
\end{equation}
where $\hat\chi^{(2)}(\mathbf r, \mathbf r', \mathbf r'')$ is the contribution to
the susceptibility quadratic in the electric induction.
Introducing Eq.~(\ref{Eq_Pol_GenL}) into Eq.~(\ref{Eq_Quad_EnQP}) we find for the interaction energy
\begin{equation}\label{Eq_Quad_EnQP1}
W^{\mathrm I}=-\frac{Q^2}{2\Omega_{\mathrm{GFM}}}\sum_{i,j}\int{d^3rd^3r' \mathbf D^{\mathrm q}_i(\mathbf r)\hat\chi(\mathbf r, \mathbf r')\mathbf D^{\mathrm q}_j(\mathbf r')}.
\end{equation}
The quadratic polarization $\mathbf P^{(2)}(\mathbf r)$ does not contribute to the interaction energy $W^{\mathrm I}$ since it
produces an odd-degree oscillating electric field $\mathbf D$.

\subsubsection{Total energy of electric field}

The total energy of electric field is given by the following expression
\begin{equation}\label{Eq_En_M}
W^{\mathrm{E}}+W^{\mathrm{I}}=Q^2R,
\end{equation}
where we introduce the notation
\begin{equation}\label{Eq_R}
R=\sum_{i,j}\int{\!\!d^3rd^3r' \mathbf D^{\mathrm q}_i(\mathbf r)\left(\frac{\delta(\mathbf r-\mathbf r')}{8\pi\Omega_{\mathrm{GFM}}}-\frac{\hat\chi(\mathbf r, \mathbf r')}{2\Omega_{\mathrm{GFM}}}\right)\mathbf D^{\mathrm q}_j(\mathbf r')}.
\end{equation}
The coefficient $R$ depends on temperature $T$ and the
external electric field $\mathbf E_0$ through the susceptibility $\hat\chi(\mathbf r, \mathbf r')$. In addition,
the coefficient $R$ also depends on the distance between the GFM film and the FE layer and on the FE thickness.

\subsection{Variational procedure}

Minimizing the total energy of the system in parameter $Q$ we obtain the equation describing the
magnitude of quadrupole
\begin{equation}\label{Eq_Quad}
2\alpha_{\mathrm Q}(Q-Q_0)+2RQ+\gamma M^2=0.
\end{equation}
This equation has the solution
\begin{equation}\label{Eq_Quad_Sol}
Q=\frac{\alpha_{\mathrm Q} Q_0-\gamma M^2/2}{\alpha_{\mathrm Q}+R}.
\end{equation}
We notice that $Q$ depends on both subsystems - the GFM film magnetization and the FE layer
polarization through coefficient $R$ leading to the coupling between the FE
polarization $P$ and the GFM magnetization $M$.

The equation describing the magnetization behaviour (up to linear in parameter $\gamma$ terms)
has the form
\begin{equation}\label{Eq_Mag_Gen}
\begin{split}
&2\tilde\alpha_{\mathrm M}\mathbf{M}+4\beta_{\mathrm{M}}M^2\mathbf{M}=\mathbf{B},\\
&\tilde\alpha_{\mathrm M}=\alpha_{\mathrm{M}}-\gamma\frac{RQ_0}{\alpha_{\mathrm Q}+R}.\\
\end{split}
\end{equation}
The magnetization ${\mathbf M}$ is parallel to the plane of the GFM film.
The magnetic field existing at the film edges is negligible due to large area of the film.
We assume that the magnetization $\mathbf M$ in Eq.~(\ref{Eq_Mag_Gen}) is uniform
because the domain wall thickness in the GFM film is much larger than the
intergrain distance and the film thickness.

The coefficient $\tilde\alpha_{\mathrm M}$ depends
on the FE state through coefficient $R$  and have some peculiarities in the vicinity of the
FE Curie point due to singularities in the susceptibility $\hat\chi(\mathbf r, \mathbf r')$.
Since the coefficient $R$ depends on the external field $\mathbf E_0$
one can control the magnetic state of the GFM film by the electric field.
The influence of the GFM film on the FE layer is finite due to electric field created by quadrupoles.

\section{FE without spatial dispersion}\label{Sec:WD}

In the absence of spatial dispersion the FE susceptibility has the form
\begin{equation}\label{Eq_Chi_DispLess}
\hat\chi(\mathbf r, \mathbf r')=\hat\chi\delta(\mathbf r-\mathbf r').
\end{equation}
Substituting this result into Eq.~(\ref{Eq_R}) we find the following result for coefficient $R$
\begin{equation}\label{Eq_R1}
\begin{split}
&R=R_0-\chi_{||}R_{||}-\chi_{\perp}R_{\perp},\\
&R_0=\frac{1}{8\pi\Omega_{\mathrm{GFM}}}\sum_{i,j}\int{\!\!d^3r \mathbf D^{\mathrm q}_{i}(\mathbf r)\mathbf D^{\mathrm q}_{j}(\mathbf r)}, \\
&R_{||}=\frac{1}{2\Omega_{\mathrm{GFM}}}\sum_{i,j}\int_{\Omega_{\mathrm{FE}}}{\!\!d^3r D^{\mathrm q}_{i(||)}(\mathbf r)D^{\mathrm q}_{j(||)}(\mathbf r)},\\
&R_{\perp}=\frac{1}{2\Omega_{\mathrm{GFM}}}\sum_{i,j}\int_{\Omega_{\mathrm{FE}}}{\!\!d^3r \mathbf D^{\mathrm q}_{i(\perp)}(\mathbf r)\mathbf D^{\mathrm q}_{j(\perp)}(\mathbf r)}.
\end{split}
\end{equation}
We assume that the FE layer has the anisotropy axis perpendicular to the layer surface. The
quantities $\chi_{||}$ and $\chi_{\perp}$ describe the longitudinal and perpendicular
susceptibility, respectively. The subscripts $||$ and $\perp$ define the longitudinal and perpendicular
components of electric induction.

To find the susceptibility in the absence of spatial dispersion we need to solve the following equation
\begin{equation}\label{Eq_Pol_DispLess}
2\alpha_{\mathrm{P}}\mathbf{P}+4\beta_{\mathrm{P}}P^2\mathbf{P}=\mathbf{E}_0+\mathbf{E},
\end{equation}
which has the solution
\begin{equation}\label{Eq_Pol_SolDispLess}
\mathbf{P}^{(1)}=\hat{\chi}\mathbf D,
\end{equation}
where
\begin{equation}\label{Eq_Perm}
\begin{split}
\chi_{||}=(2(\alpha_{\mathrm{P}}+2\pi)+12\beta_{\mathrm{P}}P_0^2)^{-1},\\
\chi_{\perp}=(2(\alpha_{\mathrm{P}}+2\pi)+4\beta_{\mathrm{P}}P_0^2)^{-1}.
\end{split}
\end{equation}
It follows from Eq.~(\ref{Eq_Perm}) that for zero external field $\mathbf E_0$ the susceptibility $\hat\chi<1/(4\pi)$.

\subsection{Influence of FE layer on the GFM film}

In this subsection we investigate the influence of FE layer on the magnetic subsystem.
In the absence of spatial dispersion of the FE, Eq.~(\ref{Eq_Mag_Gen}) has the form
\begin{equation}\label{Eq_Mag_Gen2}
\begin{split}
&2(\alpha_{\mathrm M}^*+\gamma_\perp\chi_\perp+\gamma_{||}\chi_{||})\mathbf{M}+4\beta_{\mathrm{M}}M^2\mathbf{M}=\mathbf{B},\\
\end{split}
\end{equation}
with the following coefficients
\begin{equation}\label{Eq_Mag_Gen3}
\begin{split}
&\alpha_{\mathrm M}^*=\alpha_{\mathrm{M}}-\gamma R_0Q_0/\alpha_{\mathrm Q},\\
&\gamma_{\perp}=\gamma R_\perp Q_0/\alpha_{\mathrm Q},\\
&\gamma_{||}=\gamma R_{||} Q_0/\alpha_{\mathrm Q}.
\end{split}
\end{equation}
Equation~(\ref{Eq_Mag_Gen3}) is valid for $R\ll\alpha_{\mathrm Q}$ meaning that the interaction of the GFM with
the FE layer leads to the renormalization of the constant $\alpha_{\mathrm M}$. Changing the FE susceptibility
$\hat\chi(\mathbf r, \mathbf r')$ by the external electric field one can change the FM ordering temperature.
Since the susceptibility of FE has some peculiarity in the vicinity of the FE Curie point, the
magnetic properties of the GFM film should also exhibit some peculiarities in the vicinity of the
FE Curie point.
\begin{figure}
\includegraphics[width=1\columnwidth]{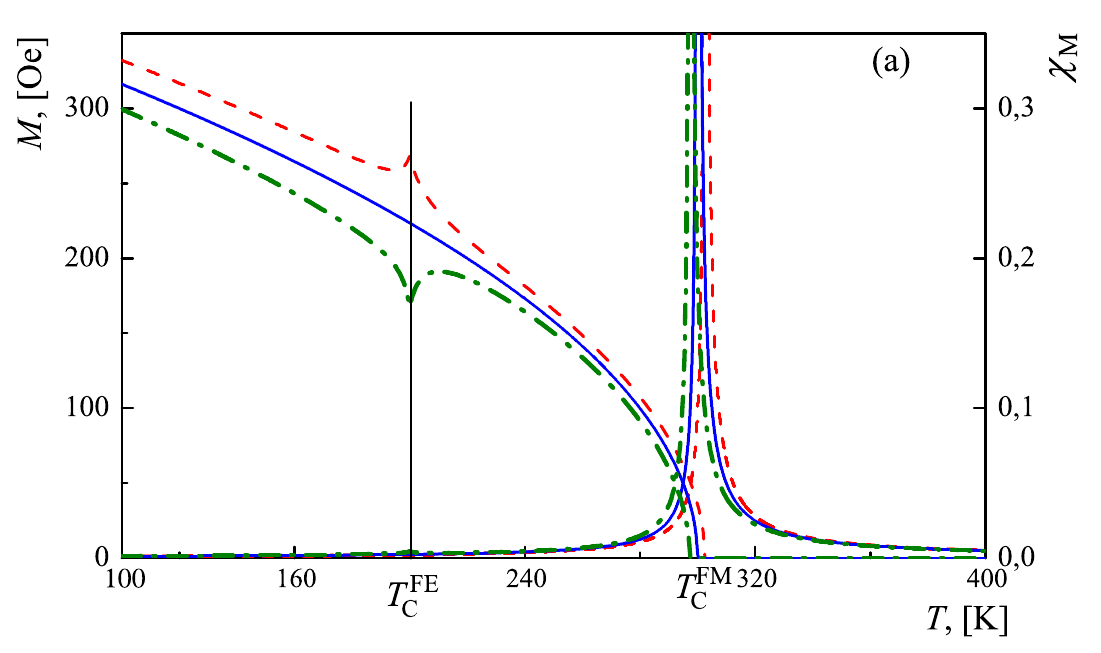}
\includegraphics[width=1\columnwidth]{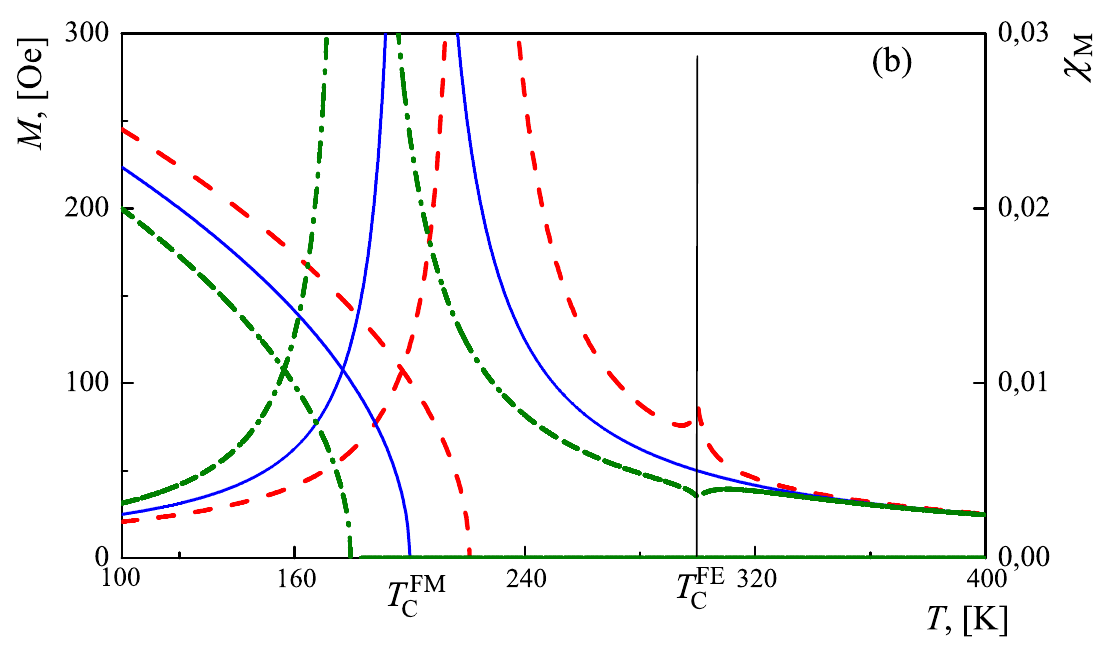}
\caption{(Color online) Magnetization $M$ and magnetic susceptibility $\chi_\mathrm M$ vs. temperature at zero magnetic and electric fields. Solid (blue) line corresponds to the absence of the FE layer. Dashed (red) line corresponds to
negative parameter $\gamma$. Dash dotted (green) line
corresponds to positive $\gamma$. $\TCFE$ and $\TCFM$ are the
ordering temperatures of the FE layer and the GFM film in the absence of mutual interaction, respectively.
(a) Limit of $\TCFE<\TCFM$.
(b) Limit of $\TCFE>\TCFM$. The interaction of FE and GFM layers leads to the appearance of peculiarities of
magnetization $M$ (panel (a)) and susceptibility $\chi_\mathrm M$ in the vicinity of the FE phase transition.}\label{Fig_WeakDispT}
\end{figure}

We assume that the coefficient $\alpha_{\mathrm M}^*=\tilde\alpha^{\mathrm{FM}}(T-\TCFM)$ in Eq.~(\ref{Eq_Mag_Gen3})
defines the position of superparamagnetic-ferromagnetic (SPM - FM) phase transition in the GFM film in the
absence of the FE layer.

The temperature dependence of magnetization and magnetic susceptibility of GFM film at zero external magnetic field is shown
in Fig.~\ref{Fig_WeakDispT}. Both limits of $\TCFE>\TCFM$ and $\TCFM>\TCFE$ are relevant
since the ordering temperature of GFM can be rather large reaching the room temperature,~\cite{Boz1972, Hel1981} and
because the FE's with the Curie point below and above the room temperature exist,~\cite{Ohigashi1983,Kitayama1981, Lee1981}.

Figure~\ref{Fig_WeakDispT}(a) shows the case $\TCFM>\TCFE$ with the following parameters:
$\alpha_\mathrm P=1(T-\TCFE)$, $\TCFE=200$ K, $\beta_\mathrm P=100$ (erg/cm$^3$)$^{-1}$ ($P_0^2=-\alpha_\mathrm P/\beta_\mathrm P$), $\alpha^*_\mathrm M=1(T-\TCFM)$ erg/(Oe$^{-2}$cm$^3$), $\TCFM=300$ K, $\beta_\mathrm M=10^{-3}$ erg/(Oe$^{-4}$cm$^3$), $\gamma_{||}=\pm300$ erg/(Oe$^{-2}$cm$^3$), $\gamma_{\perp}=\pm250$ erg/(Oe$^{-2}$cm$^3$). Since the sign of parameter  $\gamma$ is unknown we plot curves for both signs
(dashed and dash dotted lines in Fig.~\ref{Fig_WeakDispT}(a,b)). The case without FE layer is shown by solid line for comparison.

The interaction of FE and GFM layers leads to two effects: 1) The shift of the GFM film ordering
temperature which can be estimated as follows
\begin{equation}\label{Eq_TShift}
\Delta T=-\frac{\gamma_{||}\chi_{||}+\gamma_{\perp}\chi_{\perp}}{\tilde\alpha^\mathrm{FM}},
\end{equation}
where $\chi_{||,\perp}$ is taken in the vicinity of the transition temperature $\TCFM$.
The shift direction depends on the sign of interaction.

2) The peculiarity of magnetization and magnetic susceptibility
in the vicinity of the FE phase transition. The maximum deviation of magnetic susceptibility occurs at
the FE phase transition point. For $\TCFE>\TCFM$ it has the form
\begin{equation}\label{Eq_SuscDev}
\Delta \chi_\mathrm M=-\frac{\gamma_{||}\chi_{||}+\gamma_{\perp}\chi_{\perp}}{2(\tilde\alpha^\mathrm{FM}(\TCFE-\TCFM))^2}.
\end{equation}
For temperatures $\TCFM<\TCFE$ the correction is twice smaller. The change of magnetization at the FE Curie point
is
\begin{equation}\label{Eq_MagDev}
\begin{split}
\Delta (M^2)=&-\frac{\gamma_{||}\chi_{||}+\gamma_{\perp}\chi_{\perp}}{2\beta^*_\mathrm M}.
\end{split}
\end{equation}
We notice that even at the point of the FE-paraelectric phase transition the susceptibility $\hat\chi$ is finite
supporting the assumption of weak spatial dispersion.

\begin{figure}
\includegraphics[width=1\columnwidth]{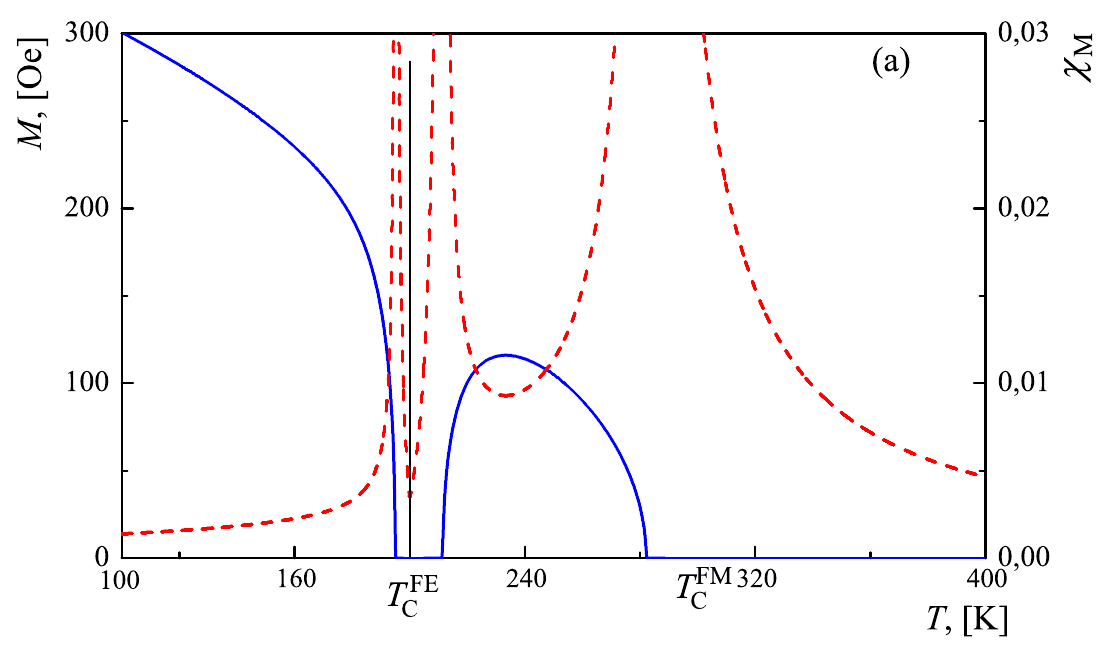}
\includegraphics[width=1\columnwidth]{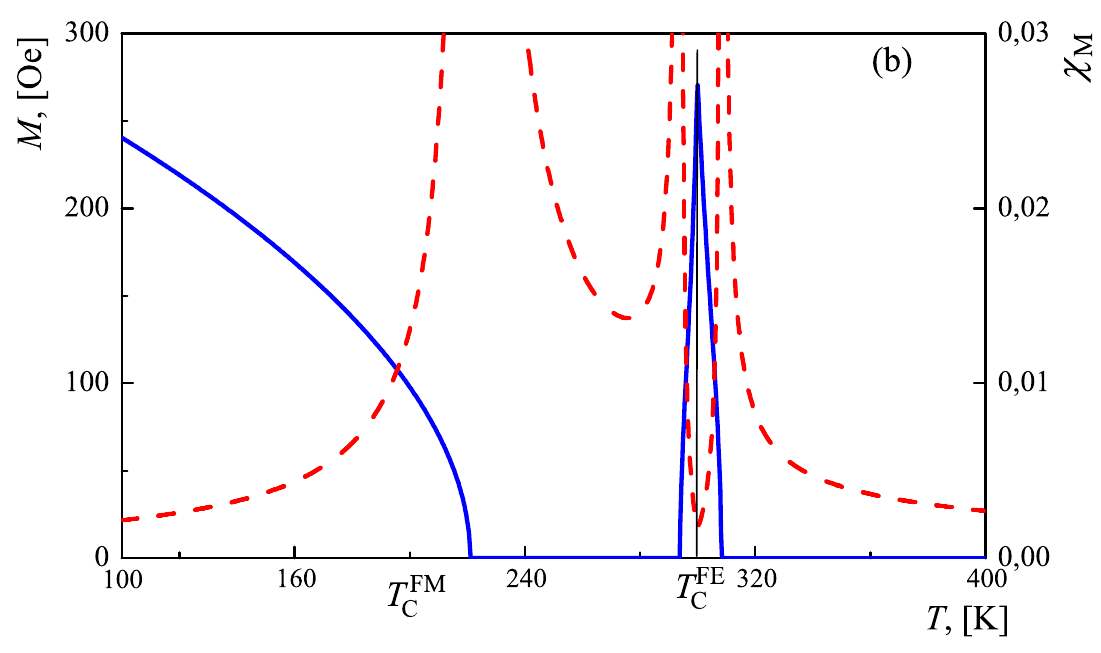}
\caption{(Color online) Magnetization $M$ (solid blue line) and magnetic susceptibility $\chi_\mathrm M$ (dashed red line) vs
temperature at zero magnetic and electric fields and strong coupling between charge fluctuations
and magnetization. (a) Limit of $\TCFE<\TCFM$, (b) Limit of $\TCFE>\TCFM$.
Two additional phase transitions occur in the vicinity of the FE phase transition.}\label{Fig_StrongEff}
\end{figure}

For large values of parameters $\gamma$ and $R_{||,\perp}$ the additional phase transitions may
occur in the vicinity of the FE phase transition, see Fig.~\ref{Fig_StrongEff}.
The curves in Fig.~\ref{Fig_StrongEff} show the ME effect discussed in Ref.~\onlinecite{Beloborodov2014_ME, Beloborodov2014_ME1} using the microscopic theory.
These curves are plotted for the same parameters as in Fig.~\ref{Fig_WeakDispT}, except $\gamma_{||}=\pm3000$ erg/(Oe$^{-2}$cm$^3$)
and $\gamma_{\perp}=\pm150$ erg/(Oe$^{-2}$cm$^3$). Sign "$+$" corresponds to Fig.~\ref{Fig_StrongEff}(a) while sign "$-$" -
to Fig.~\ref{Fig_StrongEff}(b).

\begin{figure}
\includegraphics[width=1\columnwidth]{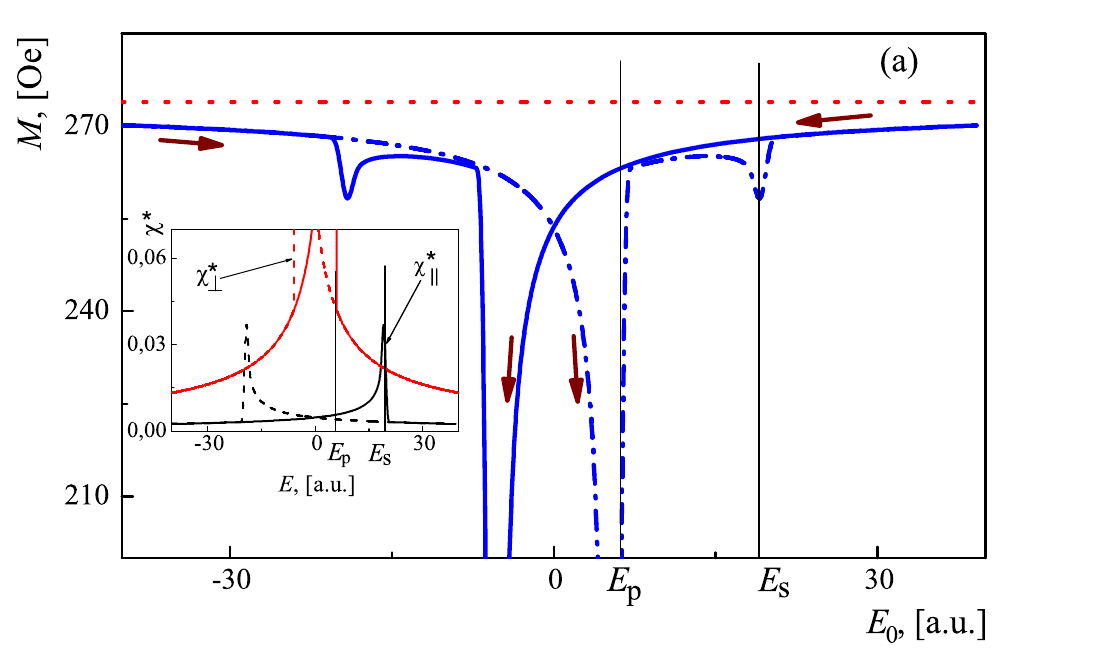}
\includegraphics[width=1\columnwidth]{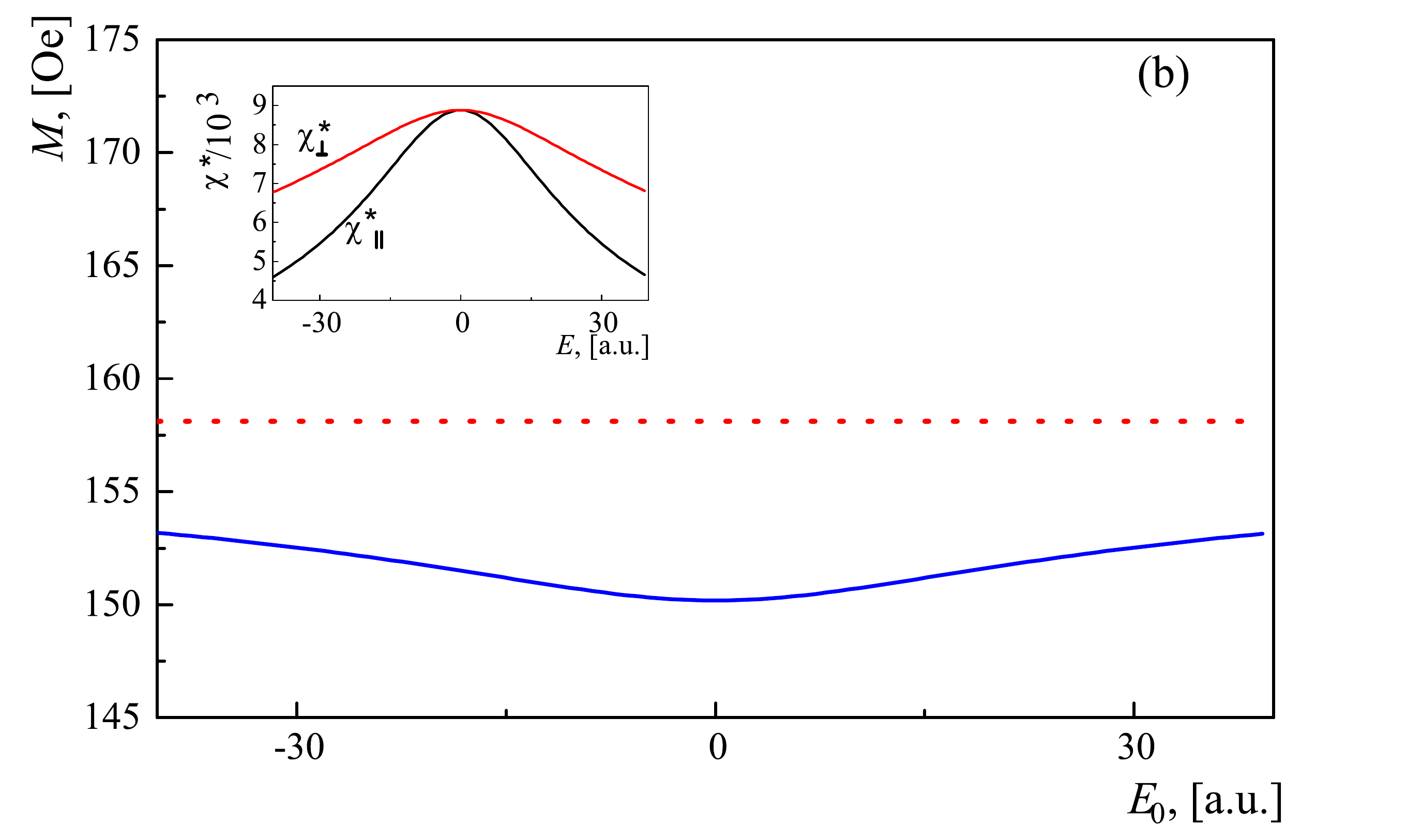}
\caption{(Color online) (a) Magnetization $M$ vs external electric field $E_0$ at zero magnetic field
for temperatures $\TCFE<\TCFM$.  Solid (blue) and dashed (blue) lines show magnetization at finite
interaction between FE layer and the GFM film. Dotted (red) line describes the non-interacting case.
Inset: dependence of $\chi_\perp$ and $\chi_{||}$ on electric field $E_0$. $E_\mathrm s$ is
the FE polarization switching field. Field $E_\mathrm p$ defines the point of $\chi_\perp$ singularity.
The hysteresis exists for temperature $T=150K < \TCFE$. (b) The same system for temperature $T=250 K > \TCFE$.}\label{Fig_MAGviaE}
\end{figure}

The dielectric susceptibility $\hat\chi$ depends on the
external electric field $E_0$. Therefore magnetic properties of the GFM film
also depend on the electric field. Figure~\ref{Fig_MAGviaE} shows the GFM magnetization
vs. external electric field $E_0$ at zero applied magnetic field.
The system parameters are the same as in Fig.~\ref{Fig_WeakDispT}, $\TCFE=200$ K and $\TCFM=300$ K.
Figure~\ref{Fig_MAGviaE}(a) is plotted for temperature $T=150K < \TCFE$. In this case the FE layer has the
spontaneous polarization. The dielectric susceptibility $\hat\chi$ strongly depends on the electric field,
see inset in Fig.~\ref{Fig_MAGviaE}. The longitudinal part $\chi_{||}$ has a peculiarity at
the point of polarization switching $\pm E_\mathrm s$. The perpendicular susceptibility $\chi_\perp$
diverges at a certain point $\pm E_\mathrm p$. Due to these peculiarities the
magnetization strongly depends on the electric field $E_0$ showing the hysteresis behavior.

Equation~(\ref{Eq_Pol_DispLess}) is not valid at points $\pm E_\mathrm p$ since the susceptibility
 $\chi_\perp$ diverges at these points and it can not be considered using perturbation theory in quadrupoles field $\mathbf D$.

\begin{figure}
\includegraphics[width=1\columnwidth]{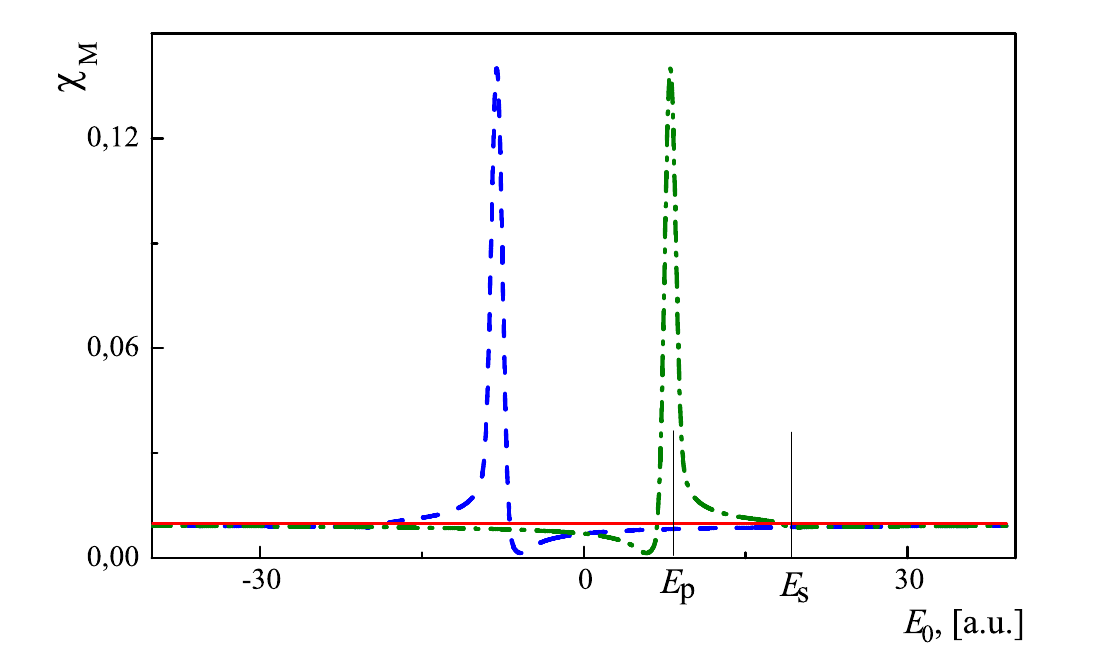}
\caption{(Color online) (a) Magnetic susceptibility $\chi_\mathrm M$ of GFM film
vs external electric field $E_0$
at zero magnetic field. The critical temperatures are $\TCFE=300$ K and $\TCFM=200$ K.
Dashed (blue) and dash-dotted (green) lines show $\chi_\mathrm M$ at
finite interaction between FE layer and the GFM film. Solid (red) line describes the non-interacting case.
$E_\mathrm s$ is the FE polarization switching field. $E_\mathrm p$ defines the singularity point of $\chi_\perp$.
The hysteresis exists for temperature $T=250 K < \TCFE$.}\label{Fig_KSIMviaE}
\end{figure}

The FE layer influences the magnetic susceptibility $\chi_\mathrm M$ for temperatures $\TCFM < \TCFE$.
Figure~\ref{Fig_KSIMviaE} shows the magnetic susceptibility  $\chi_\mathrm M$
vs. electric field $E_0$ for temperatures $\TCFM<T<\TCFE$.
In this temperature widow the susceptibility has hysteresis.

\subsection{Influence of GFM film on the FE layer}

In this subsection we investigate the influence of magnetic subsystem on the FE layer.
The correction to the polarization $P$ quadratic in the electric induction $\mathbf D$ has the form
\begin{equation}\label{Eq_Pol_quad}
\mathbf P^{(2)}=-4\beta_\mathrm P((\hat\chi\mathbf D)^2\hat\chi\mathbf P_0+2(\mathbf P_0\hat\chi\mathbf D)\hat\chi\hat\chi\mathbf D).
\end{equation}
The correction $\mathbf P^{(2)}$ averaged over the FE volume is parallel to the polarization
$\mathbf P_0$
\begin{equation}\label{Eq_Pol_quad2}
\langle\mathbf P^{(2)}\rangle=4Q^2\beta_\mathrm P \mathbf P_0\chi_\perp(3(\chi_\perp)^2 R^*_\perp+(\chi_{||})^2R^*_{||}),
\end{equation}
where $R^*_{\perp,||}=\Omega_{\mathrm{GFM}}R_{\perp,||}/\Omega_\mathrm{FE}$. Using Eq.~(\ref{Eq_Quad_Sol}) for
parameter $Q$ we find
\begin{equation}\label{Eq_Pol2_Sol}
\langle\mathbf P^{(2)}\rangle=\frac{4}{h}\left(Q_0^2-\frac{\gamma M^2}{\alpha_{\mathrm Q}}\right)\beta_\mathrm P \mathbf P_0\chi_\perp(3\chi_\perp^2 R^*_\perp+\chi_{||}^2R^*_{||}).
\end{equation}
For temperatures $T > \TCFM$ the correction $\mathbf P^{(2)}$ in the presence of external magnetic field
behaves as $\mathbf P^{(2)} \sim \chi_\mathrm M^2B_{\mathrm{ext}}^2$, while for temperatures $ T < \TCFM$ it has a hysteresis
dependence on the magnetic field, $B_{\mathrm{ext}}$.

\begin{figure}
\includegraphics[width=1\columnwidth]{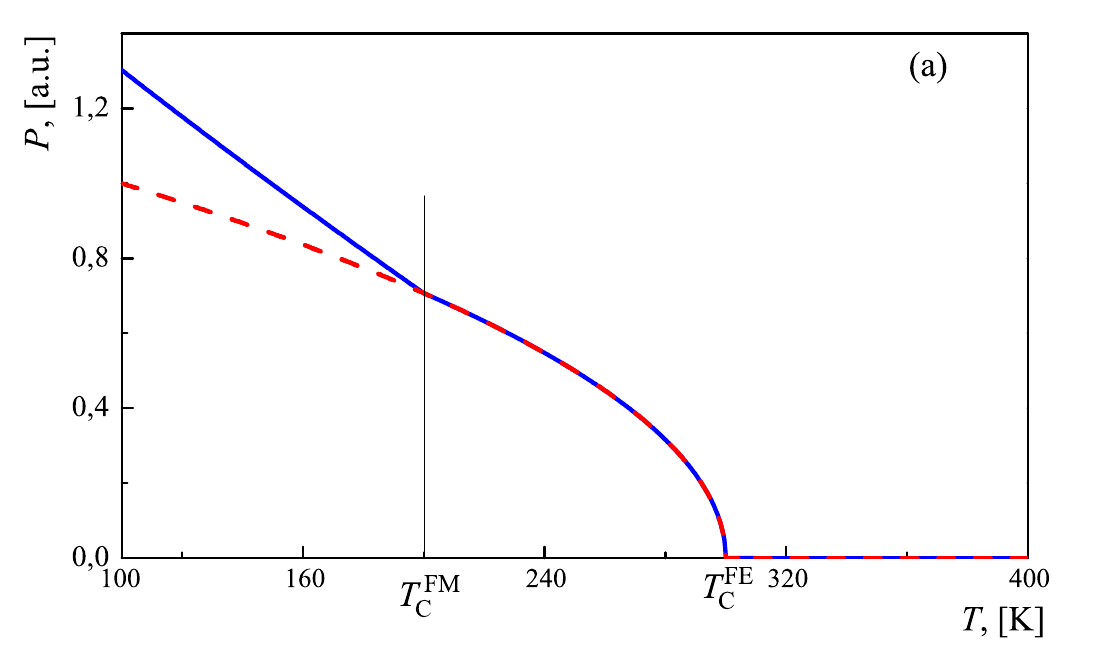}
\includegraphics[width=1\columnwidth]{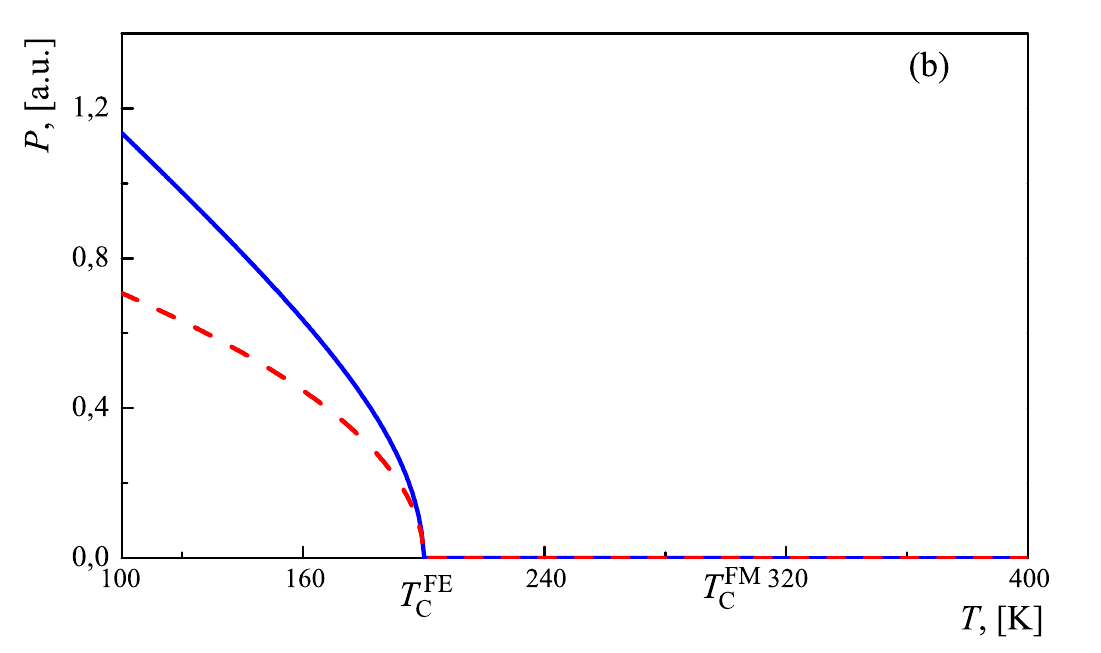}
\caption{(Color online) Average polarization $P$ of the FE layer along the $z$ direction vs temperature $T$ at
zero external magnetic and electric fields. Solid (blue) line describes the case of finite interaction of FE layer
with GFM film. Dashed (red) line corresponds to the non-interacting case. (a) Limit of $\TCFE>\TCFM$ (b) Limit of $\TCFE<\TCFM$.}\label{Fig_P_T_WD}
\end{figure}

The temperature dependence of the FE layer polarization is shown
in Fig.~\ref{Fig_P_T_WD}. The curves are plotted for the following set of parameters: $\gamma_\perp=3\cdot10^{-5}$ erg(cm$^{-3}$ Oe$^{-2}$), $\gamma_{||}=2\cdot10^{-5}$ erg(cm$^{-3}$ Oe$^{-2}$). Figure~\ref{Fig_P_T_WD}(a) corresponds to temperatures $\TCFE>\TCFM$, while
Fig.~\ref{Fig_P_T_WD}(b) is plotted for $\TCFE<\TCFM$.

\subsection{Dependence of Magneto-Electric coupling on the system parameters}\label{Sec_R}

We use the Ewald approach to calculate the electric field of two dimensional periodic lattice of quadrupoles.~\cite{Avoird1981}
The field produce by this lattice is periodic in the (x,y)-plane and decays along the z direction.
The spatial Fourier harmonics of the field are given by~\cite{}
\begin{equation}\label{Eq_Field}
\begin{split}
&E_{x,y}(\mathbf k_\perp,z)=i\sqrt{\frac{4\pi}{5}}\frac{\pi}{L_\mathrm g^2}(k_{x,y}) k_\perp E^{-k_\perp z} \times \\&\times \left(\frac{2\cos(2\phi_\perp)}{\sqrt{6}}\left[Q_2^{(2)}+\tilde Q_2^{(2)}e^{i\mathbf k_\perp\mathbf s}\right]+Q_0^{(2)}+\tilde Q_0^{(2)}e^{i\mathbf k_\perp\mathbf s}\right),\\
&E_{z}(\mathbf k_\perp,z)=-\sqrt{\frac{4\pi}{5}}\frac{\pi}{L_\mathrm g^2}k^2_\perp E^{-k_\perp z} \times \\&\times \left(\frac{2\cos(2\phi_\perp)}{\sqrt{6}}\left[Q_2^{(2)}+\tilde Q_2^{(2)}e^{i\mathbf k_\perp\mathbf s}\right]+Q_0^{(2)}+\tilde Q_0^{(2)}e^{i\mathbf k_\perp\mathbf s}\right),
\end{split}
\end{equation}
where $\mathbf k_\perp=(k_x,k_y,0)$, $\phi_\perp=\arctan(k_x/k_y)$. The wave vector $\mathbf k_\perp$ has
the discrete values $\mathbf k^{n,m}_\perp=(2\pi n/L_\mathrm g,2\pi m/L_\mathrm g, 0)$.  There are
two quadrupoles, $\hat Q^{1}$ and $\hat Q^{2}$, in a unit cell.
The vector $\mathbf s$ defines the shift of these dipoles, $\mathbf s=(\pi/L_\mathrm g,\pi/L_\mathrm g, 0)$.
The parameters $Q^{(2)}_i$ and $\tilde Q^{(2)}_i$ are related to $Q$ as follows
$Q^{(2)}_0=-Q$, $\tilde Q^{(2)}_0=-Q$, $Q^{(2)}_2=-3Q/(2\sqrt{6})$, $\tilde Q^{(2)}_2=3Q/(2\sqrt{6})$.

The magnitude of spatial Fourier harmonic in Eq.~(\ref{Eq_Field})
decreases exponentially with increasing the vector $k_\perp$.
Therefore even for $z=L_{\mathrm g}$ we can neglect all harmonics except
the four harmonics nearest to zero, $(\pm 2\pi/L_\mathrm g,0,0)$ and $(0,\pm 2\pi/L_\mathrm g,0)$.
Using Eq.~(\ref{Eq_Field}) we obtain
\begin{equation}\label{Eq_Field1}
\begin{split}
&E_{x}(\pm 2\pi/L_\mathrm g,0,z)=\pm i\sqrt{\frac{4\pi}{5}}\frac{4\pi^3}{L_\mathrm g^4} e^{-2\pi z/L_\mathrm g} \times \\&\times \left(\frac{2}{\sqrt{6}}\left[Q_2^{(2)}+\tilde Q_2^{(2)}e^{i\mathbf k_\perp\mathbf s}\right]+Q_0^{(2)}+\tilde Q_0^{(2)}e^{i\mathbf k_\perp\mathbf s}\right),\\
&E_{y}(\pm 2\pi/L_\mathrm g,0,z)=0,\\
&E_{z}(\pm 2\pi/L_\mathrm g,0,z)=-\sqrt{\frac{4\pi}{5}}\frac{4\pi^3}{L_\mathrm g^4} e^{-2\pi z/L_\mathrm g}  \times \\&\times \left(\frac{2}{\sqrt{6}}\left[Q_2^{(2)}+\tilde Q_2^{(2)}e^{i\mathbf k_\perp\mathbf s}\right]+Q_0^{(2)}+\tilde Q_0^{(2)}e^{i\mathbf k_\perp\mathbf s}\right),\\
&E_{x}(0,\pm 2\pi/L_\mathrm g,z)=0,\\
&E_{y}(0,\pm 2\pi/L_\mathrm g,z)=\pm i\sqrt{\frac{4\pi}{5}}\frac{4\pi^3}{L_\mathrm g^4} e^{-2\pi z/L_\mathrm g} \times \\&\times \left(-\frac{2}{\sqrt{6}}\left[Q_2^{(2)}+\tilde Q_2^{(2)}e^{i\mathbf k_\perp\mathbf s}\right]+Q_0^{(2)}+\tilde Q_0^{(2)}e^{i\mathbf k_\perp\mathbf s}\right),\\
&E_{z}(0,\pm 2\pi/L_\mathrm g,z)=-\sqrt{\frac{4\pi}{5}}\frac{4\pi^3}{L_\mathrm g^4} e^{-2\pi z/L_\mathrm g}  \times \\&\times \left(-\frac{2}{\sqrt{6}}\left[Q_2^{(2)}+\tilde Q_2^{(2)}e^{i\mathbf k_\perp\mathbf s}\right]+Q_0^{(2)}+\tilde Q_0^{(2)}e^{i\mathbf k_\perp\mathbf s}\right).
\end{split}
\end{equation}
The amplitude of electric field oscillations decays with distance
as $e^{-2\pi z/L_\mathrm g}$. The parameter $R$ is averaged over the volume of the FE ($d<z<h+d$).
Using Eq.~(\ref{Eq_Field1}) we find
\begin{equation}
\label{Eq_R_Behav}
R_{||,\perp}\sim e^{-4\pi d/L_\mathrm g}\left(1-e^{-4\pi h/L_\mathrm g}\right).
\end{equation}
The magneto-electric coupling exponentially decays with increasing the distance between the GFM
film and the FE layer with the characteristic decay length being the intergrain distance, $L_\mathrm g$.

The coefficients $R$ saturates with increasing the FE thickness $h$ due to the exponential
decay of the electric field with distance $d$. The saturation occurs for thickness's $h$ larger than
the intergrain distance $L_\mathrm g$ leading to weak influence of the GFM film on the FE layer.

\section{FE with strong spatial dispersion}\label{Sec:SD}

\subsection{Influence of FE layer on the GFM film}

The coupling between the FE layer and the GFM film depends on the parameter $R$, see Eq.~(\ref{Eq_R}). Above
we discussed the case of FE without spatial dispersion meaning that the FE
response $\hat\chi(\mathbf r, \mathbf r')$ is local. In the opposite case, of strong spatial dispersion
we can consider $\hat\chi(\mathbf r, \mathbf r')= const $, being independent of coordinates.
In this case Eq.~(\ref{Eq_R}) has the form
\begin{equation}\label{Eq_R_StrDisp}
R=R_0-\left(\sum_{i}\int{\!\!\mathbf D^{\mathrm q}_i(\mathbf r)d^3r}\right)\frac{\hat\chi}{2\Omega_{\mathrm{GFM}}}\left(\sum_{j}\int{\!\!\mathbf D^{\mathrm q}_j(\mathbf r)d^3r}\right).
\end{equation}
The average field created by the ensemble of quadrupoles is zero.
Therefore, for strong spatial dispersion the FE layer and the GFM film
are decoupled since the parameter $R\to R_0$. Thus, below we consider the quantity $R$ with large
but finite spatial dispersion.

The linear response of the FE layer is described by the following equation
\begin{equation}\label{Eq_Pol_Disp}
-\delta_{\mathrm{P}}\Delta\mathbf{P}^{(1)}+\hat{\chi}^{-1}\mathbf{P}^{(1)}=\mathbf D.
\end{equation}
This equation differs from Eq.~(\ref{Eq_Pol_DispLess}) by the term with spatial derivatives
responsible for dispersion. We use the following boundary condition for polarization,
$\mathbf (P^{(1)})'_z=0|_{z=h,h+d}$, with $h$ and $h+d$ being
the boundary position of the FE layer,~\cite{Frid2006rev,Frid2010rev,Tilley2001}.

It was shown in Sec.~\ref{Sec_R} that the electric field $\mathbf D$ produced by
the lattice of quadrupoles is periodic in the (x,y) plane and decays in the z-direction.
For distances $|z| > L_\mathrm g$ away from the GFM film the field has (x,y) spatial Fourier
harmonics with only $|\mathbf k_\perp|=2\pi/ L_\mathrm g$ and the decay length $k_\mathrm d=2\pi/ L_\mathrm g$.
Such a field can be considered as a wave with zero wavevector $|\mathbf k|^2=|\mathbf k_\perp|^2-k_\mathrm d^2=0$.
Therefore the partial solution of Eq.~(\ref{Eq_Pol_Disp}) has the form
\begin{equation}\label{Eq_Pol_PartDisp}
\mathbf{P}^{(1)}_{\mathrm p}=\hat{\chi}\mathbf D.
\end{equation}
And the uniform solution has the form
\begin{equation}\label{Eq_Pol_PartDisp}
\mathbf{P}^{(1)}_{\mathrm u}=\mathbf C_1e^{-\hat\lambda z}+\mathbf C_2e^{\hat\lambda z},
\end{equation}
where the vectors $\mathbf C_1$ and $\mathbf C_2$ depend on the $x$ and $y$ coordinates
similar to the electric field $\mathbf D$.
\begin{equation}\label{Eq_Lambda}
\hat{\lambda} = \sqrt{\mathbf k_{\perp}^2+\hat{\chi}^{-1}/\delta_{\mathbf P}}.
\end{equation}
$\hat{\lambda}$ is the tensor. The appropriate components of tensor $\hat{\chi}^{-1}$
should be used for each vector component $\mathbf C_{1,2}$.
Using the boundary conditions we find the coefficients $\mathbf C_{1,2}$
\begin{equation}\label{Eq_CoefC}
\begin{split}
&\mathbf C_1=\frac{k_\perp\hat\chi\tilde{\mathbf D}}{\hat\lambda(e^{\hat\lambda h}-e^{-\hat\lambda h})}e^{-k_\perp d -\hat\lambda d}(e^{-k_\perp h}-e^{-\hat\lambda h}),\\
&\mathbf C_2=\frac{k_\perp\hat\chi\tilde{\mathbf D}}{\hat\lambda(e^{\hat\lambda h}-e^{-\hat\lambda h})}e^{-k_\perp d +\hat\lambda d}(e^{-k_\perp h}-e^{\hat\lambda h}).
\end{split}
\end{equation}
Here $\tilde{\mathbf D}=e^{k_\perp z} \mathbf D$. $\tilde{\mathbf D}$ depends on the
coordinates $x$ and $y$ only, since ${\mathbf D}\sim e^{-k_\perp z}$.
For strong spatial dispersion and thick FE layer the linear polarization has the form
\begin{equation}\label{Eq_Pol_Sol_thick_Disp}
\begin{split}
&\mathbf P^{(1)}=\mathbf P^{(1)}_{\mathrm p}+\mathbf P^{(1)}_{\mathrm u}=\frac{\mathbf D}{2\delta_\mathrm P k_\perp^2}\times \\
&\times \left((z-d)k_\perp+1-\frac{3+3(z-d)k_\perp+(z-d)^2k_\perp^2}{4\chi\delta_\mathrm P k_\perp^2}\right).
\end{split}
\end{equation}
The characteristic length scale for coefficients $R_{||,\perp}$ is the distance between two centres of
neighbouring grains $L_\mathrm g$. This is the consequences of the fact that the electric induction $\mathbf D$ in the FE layer decays exponentially.
For estimates we use $z-d\approx L_\mathrm g$ and $(z-d)k_\perp\approx 1$. Thus, we find for polarization
\begin{equation}\label{Eq_Pol_Sol_thick_Disp1}
\mathbf P^{(1)}\approx\frac{\mathbf D}{\delta_\mathrm P k_\perp^2}\left(1-\frac{7}{8\hat\chi\delta_\mathrm P k_\perp^2}\right).
\end{equation}

Using Eq.~(\ref{Eq_Pol_Sol_thick_Disp1}) we calculate the coefficient $R$
\begin{equation}\label{Eq_R_thick_Disp1}
\begin{split}
&R= \tilde R_0-\frac{\tilde R_{||}}{\chi_{||}}-\frac{\tilde R_{\perp}}{\chi_\perp},\\
&\tilde R_0=R_0\left(1+\frac{L^2_\mathrm g}{\delta_\mathrm P 4\pi^2}\right),\\
&\tilde R_{||}=-\frac{7L^4_\mathrm g R_{||}}{8\chi_{||}\delta^2_\mathrm P (4\pi^2)^2},\\
&\tilde R_{\perp}=-\frac{7L^4_\mathrm g R_{\perp}}{8\chi_{\perp}\delta^2_\mathrm P (4\pi^2)^2}.
\end{split}
\end{equation}
The coefficient $R_{||,\perp}$ is calculated using Eq.~(\ref{Eq_R1}) with electric field given by Eq.~(\ref{Eq_Field1}).
It follows that the influence of the FE layer on the GFM film is suppressed for strong
spatial dispersion by the factor $L_\mathrm g^2/\delta_\mathrm P$. The coefficient
$\tilde R_{||,\perp}$ have the opposite sign to the coefficient $R_{||,\perp}$.

The equation for magnetization has the form
\begin{equation}\label{Eq_Mag_Gen2SD}
\begin{split}
2\left(\tilde\alpha_{\mathrm M}^*+\frac{\tilde\gamma_\perp}{\chi_\perp}+\frac{\tilde\gamma_{||}}{\chi_{||}}\right)\mathbf{M}+
4\tilde\beta_{\mathrm{M}}M^2\mathbf{M}=\mathbf{B},\\
\end{split}
\end{equation}
with the following coefficients
\begin{equation}\label{Eq_Mag_Gen3SD}
\begin{split}
&\tilde\alpha_{\mathrm M}^*=\alpha_{\mathrm{M}}-\gamma \tilde R_0Q_0/\alpha_{\mathrm Q},\\
&\tilde\gamma_{\perp}=\gamma \tilde R_\perp Q_0/\alpha_{\mathrm Q},\\
&\tilde\gamma_{||}=\gamma \tilde R_{||} Q_0/\alpha_{\mathrm Q}.
\end{split}
\end{equation}
In contrast to the weak dispersion case, here the susceptibility $\chi$ is present in
the denominator leading to a different dependence of magnetization on temperature and electric field.

\begin{figure}
\includegraphics[width=1\columnwidth]{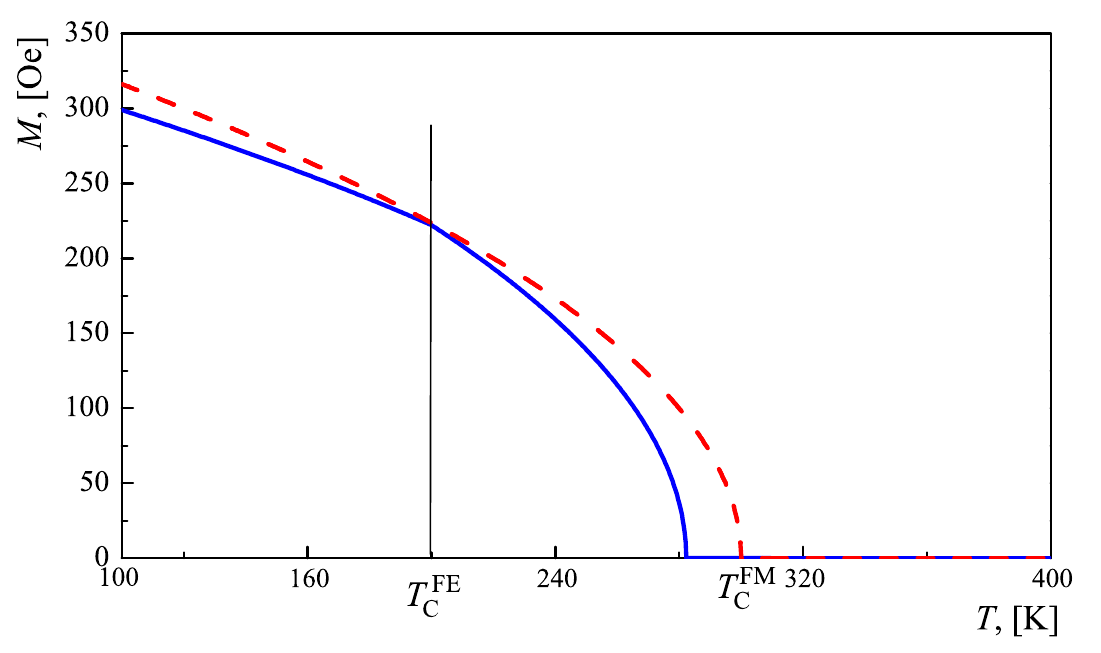}
\caption{(Color online) Magnetization $M$ vs temperature $T$ for strong spatial dispersion and zero
external magnetic and electric fields. Solid (blue) line describes the case of finite interaction of the
FE layer with the GFM film while the dashed (red) line corresponds to the non-interacting case.}\label{Fig_SD_M_T}
\end{figure}

Figure~\ref{Fig_SD_M_T} shows the magnetization $M$ behavior in the vicinity of the critical temperature $\TCFE$
for strong spatial dispersion. All parameters for GFM film and the FE layer are the same as before. The parameter
$\delta_\mathrm P$ was chosen such that $\tilde\gamma_{\perp}=0.05$ erg/(Oe$^{-2}$cm$^3$) and $\tilde\gamma_{||}=0.05$ erg/(Oe$^{-2}$cm$^3$). The influence of the
FE layer on the magnetization in the case of strong dispersion is the opposite to the
case of weak dispersion: It is small in the vicinity of the
FE-PE phase transition and increases with increasing the distance from the critical temperature $\TCFE$.
In general, increasing the difference $|T-\TCFE|$ one can study the crossover from strong to weak dispersion.
Thus, the dependence of magnetization on temperature can be considered as a combination of Figs.~\ref{Fig_WeakDispT}
and~\ref{Fig_SD_M_T}. The crossover temperature between two regimes depends on the system parameters.

\begin{figure}
\includegraphics[width=1\columnwidth]{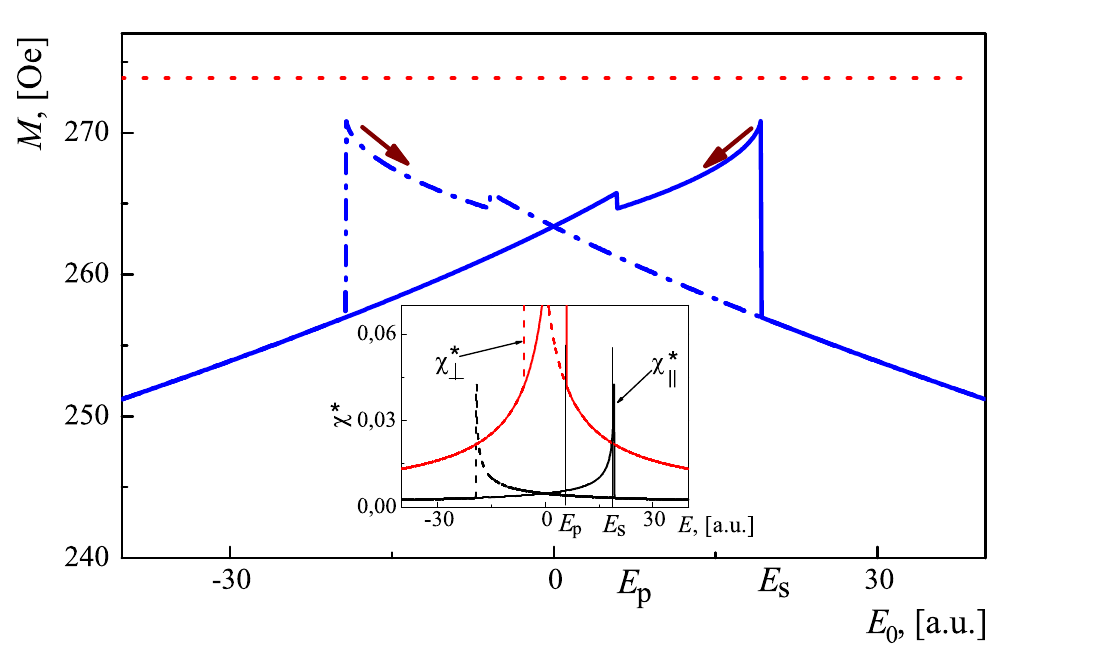}
\caption{(Color online) Magnetization $M$ vs external electric field $E_0$ for strong spatial dispersion and
zero external magnetic field. Dash-dotted (blue) lines describe the case of finite interaction of FE layer
with GFM film. Dotted (red) line corresponds to the non-interacting case. The plots are shown for the following
sets of parameters: $T=150$ K, $\TCFE=200$ K, $\TCFM=300$ K, $\tilde\gamma_{\perp}=0.05$ erg/(Oe$^{-2}$cm$^3$) and $\tilde\gamma_{||}=0.05$ erg/(Oe$^{-2}$cm$^3$). Parameters $\tilde\xi_{\perp,||}$ are negligibly small. Inset:
Susceptibility $\hat\chi$ vs electric field $E_0$. }\label{Fig_SD_M_E}
\end{figure}

The magnetization $M$ vs. external electric field $E_0$ is shown in Fig.~\ref{Fig_SD_M_E} for $\TCFM > \TCFE$
and fixed temperature $T=150$ K. In contrast to the limit of weak dispersion,
where magnetization $M$ has some peculiarities at fields $\pm E_p$, the magnetization in this case has the peculiar
points at fields $E_0=\pm E_s$. This is the consequence of the fact that the susceptibility $\hat\chi$ is
present in the denominator.

For thin FE layer the polarization is given by
\begin{equation}\label{Eq_Pol_Sol_thin_Disp}
\mathbf P^{(1)}=\mathbf P^{(1)}_{\mathrm p}+\mathbf P^{(1)}_{\mathrm u}=\frac{\mathbf D}{\delta_\mathrm P  k_\perp^2}\left(1-\frac{1}{\hat\chi\delta_\mathrm P k_\perp^2}\right).
\end{equation}
This polarization produces similar behavior of magnetization as a function of temperature and electric
field with slightly modified coefficients. For thin FE film the
coefficients $\tilde R_{\perp,||}$ are linearly depend on the FE thickness, $h$.

\subsection{Influence of GFM film on the FE layer}

In this subsection we investigate the influence of GFM film on the FE layer in the case of strong dispersion.
The equation describing the part of polarization quadratic in the electric induction has the form
\begin{equation}\label{Eq_Pol2_SD}
\begin{split}
-\delta_\mathrm P \Delta \mathbf P^{(2)}+(\hat\chi)^{-1}\mathbf P^{(2)}=-4\beta_\mathrm P&(2(\mathbf P_0 \mathbf P^{(1)})\mathbf P^{(1)}+\\+&(\mathbf P^{(1)})^2\mathbf P_0).
\end{split}
\end{equation}
To solve Eq.~(\ref{Eq_Pol2_SD}) we use the same boundary conditions as we used before for $\mathbf P^{(1)}$.
We are interested in average polarization $\mathbf P^{(2)}$ appearing
due to nonlinear response. Only the average z-component of $\mathbf P^{(2)}$ is non-zero.
$P_z^{(2)}$ has a contribution with $\mathbf k_\perp=0$. For this component we have
\begin{equation}\label{Eq_Pol2_SD2}
\begin{split}
\delta_\mathrm P \frac{\partial^2}{\partial z^2} &P_z^{(2)}-(\hat\chi)^{-1}P_z^{(2)}= \\ &=\frac{4\beta_\mathrm P P_0}{(\delta_\mathrm P k_\perp^2)^2}(3\langle D_z^2\rangle_{x,y}+\langle \mathbf D_\perp^2\rangle_{x,y}).
\end{split}
\end{equation}
Here the notation $\langle\rangle_{x,y}$ stands for averaging over the (x,y) plane.
The field $\mathbf D^2$ decays with distance as $e^{-2k_\perp z}$, where
$k_\perp=2\pi/L_\mathrm g$. Therefore the partial solution of Eq.~(\ref{Eq_Pol2_SD2}) has the form
\begin{equation}\label{Eq_Pol_SD_Sol_Par}
P_{z(\mathrm p)}^{(2)}=\frac{\beta_\mathrm P P_0}{(\delta_\mathrm P k_\perp^2)^3}(3\langle D_z^2\rangle_{x,y}+\langle \mathbf D_\perp^2\rangle_{x,y}).
\end{equation}
We neglect the term with the susceptibility $(\hat\chi)^{-1}$ in Eq.~(\ref{Eq_Pol2_SD2}).
The uniform solution for $k_\perp=0$ has the form
\begin{equation}\label{Eq_Pol_SD_Sol_Un}
P_{z(\mathrm u)}^{(2)}=C^\mathrm z_1e^{\lambda^*z}+C^\mathrm z_2e^{-\lambda^*z},
\end{equation}
where $\lambda^*=\sqrt{1/(\chi_{||}\delta_\mathrm P)}$.

Using the boundary condition we find that $C^z_i\sim1/(\delta_\mathrm P k_\perp^2)^{5/2}$ with $k_\perp=2\pi/L_\mathrm g$.
Therefore the average polarization $P_z^{(2)}$ decays with increasing the spatial dispersion as $(\delta_\mathrm P k_\perp^2)^{-5/2}$.
For strong dispersion the correction $P_z^{(2)}$ is also quadratic
in parameter $Q$ leading to the same behavior of average polarization on the magnetic field
as in the case of weak dispersion. However, the influence of the GFM film on the FE layer is suppressed due to
spatial dispersion.

\section{Microscopic model of coupling between quadrupole moment and magnetization}\label{Sec:Micro}

In Ref.~\onlinecite{Beloborodov2014_ME} we developed the model describing the coupling
between electric and magnetic degrees of freedom in the GMF.
The coupling mechanism is based on the interplay of intergrain
exchange coupling, Coulomb blockade and screening of electric field by the FE polarization.
In this model the exchange interaction of two neighbouring grains appears due to the overlap of
electron wave functions in the space between the grains, see Fig.~\ref{Fig_Quadr}(b)
\begin{equation}
\label{Eq_exch}
J\propto\sum\int\Psi^{*}_1(\mathbf{r}_2)\Psi_2^{*}(\mathbf{r}_1)U_c(\mathbf r_1-\mathbf r_2)\Psi_1(\mathbf{r}_1)\Psi_2(\mathbf{r}_2)d\mathbf r_1d\mathbf r_2.
\end{equation}
Here $\Psi_{1,2}$ is the spatial part of the electron wave function located in the
first (second) grain; $U_c$ is the Coulomb interaction of electrons located in different grains.
Summation is over the different electron pairs in the grains.

\begin{equation}
\label{WaveFunc}
\Psi_{1,2}(\mathbf{r}) = A \left\{\begin{array}{l}e^{-\frac{a}{\xi}},~~|\mathbf{r}\pm\mathbf{L}_{\mathrm{g}}/2|<a, \\e^{-\frac{|\mathbf{r}\pm\mathbf{L}_{\mathrm{g}}/2|}{\xi}},~~|\mathbf{r}\pm\mathbf{L}_{\mathrm{g}}/2|>a. \end{array} \right.
\end{equation}
Here $A$ is the normalization constant and $L_{\mathrm{g}}$ is the
distance between two grain centres. $\xi$ is the electron localization length.
It depends on the dielectric permittivity of the FE leading to the
strong influence of the FE state on the intergrain exchange interaction and consequently on the magnetic state of granular film,~\cite{Beloborodov2014_ME}.

For small localization length, $\xi < \min (a, L_\mathrm g) $, the exchange interaction
has the form $J\sim\xi^2e^{-\kappa L_\mathrm g /\xi}$, where $\kappa$ is a positive number
of order one. At equilibrium, without FE, this expression can be linearized
in $\xi$ around $\xi_0$, $J=J_0+(\xi-\xi_0)\tilde\gamma$, where $\xi_0$ is the localization length
in the absence of FE layer.
Changing the localization length $\xi$ one can control the exchange interaction and thus the
magnetic state of granular film.

For small localization length, $\xi\ll a$, one can calculate the quadrupole moment of two electrons
between the grains, $Q_{xx}\approx \xi e(3a/5-9L_\mathrm g/16)$, $Q_{yy}=Q_{zz}=-1/2Q_{xx}$, $Q=Q_{xx}+Q_{yy}=1/2Q_{xx}$.
Calculating $Q_{xx}$ we assumed that positively charged ions are located inside the grains and we
averaged over the region between the centres of two grains, $-L_\mathrm g/2<z<L_\mathrm g/2$, see Fig.~\ref{Fig_Quadr}(b).
Thus, the  quadrupole moment $Q$ is a linear function of localization
length $\xi$ and therefore the exchange interaction can be written as $J=J_0+(Q-Q_0)\gamma$.

\section{discussion}\label{Sec:Disc}

In this section we discuss the validity of our model.
The real granular films can not be described by the regular lattice since
materials have always some degree of disorder. The quadrupole moments fluctuate in space, magnitude, and
orientation due to this randomness. However, the presence of disorder does not change
qualitatively our main results. In particular,
the electric field produced by the GFM film decays exponentially with distance leading to
the same results. The coupling between the GFM film and the FE layer decreases with increasing the
spatial dispersion of the FE layer. This effect is suppressed for FEs with domain wall thickness
exceeding the average intergrain distance. For strongly disordered films one can use a
continuous spatial distribution of quadrupole moments.

For multilayer system of grains only the nearest layer to the FE substrate will interact with the
FE due to the exponential decay of coupling with distance.

In our consideration we used a certain type of boundary conditions for FE polarization,
with polarization derivatives being zero at the interface. In general, one can use the following
combination for boundary conditions, $\zeta_1 P+\zeta_2 (P)'_z=0$. It does not change qualitatively
our results.

\section{Conclusion}

We described the coupling between the FE polarization  and magnetization of GFM film using a
phenomenological model of combined multiferroic system consisting of granular ferromagnet film placed above
the FE layer. We showed that the coupling is due to the presence of oscillating in space electric charges in the GFM film.
On one hand these charges interact with the FE layer via Coulomb interaction. On the other hand they are
coupled with the magnetization leading to the mutual influence of the FE polarization and the GFM
film magnetization even for space separated FE layer and the GFM film. This model allows to study the importance
of spatial dispersion of FE polarization and to understand the influence of GFM film on the FE polarization.

We studied the temperature and electric field dependence of magnetization and magnetic susceptibility
of GFM film for weak and strong spatial dispersion of the FE layer.
We calculated the electric polarization as a function of temperature and magnetic field and
investigated the influence of the FE state on the magnetization and magnetic susceptibility and vice versa.
The effect of mutual influence decreases with increasing the spatial dispersion of the FE layer.
For weak dispersion the strongest coupling occurs in the vicinity of the FE-PE phase transition.
For strong dispersion the situation is the opposite. We showed that for temperatures $T < \TCFE$ the
magnetization has hysteresis as a function of electric field. For strong coupling the interaction of the FE layer
and the GFM film leads to the appearance of an additional magnetic phase transition. Below the ordering
temperature of GFM film the
FE polarization has hysteresis as a function of magnetic field.

We studied the behavior of magneto-electric coupling as a function of distance between the FE layer
and the GFM film.
We showed that for large distances the coupling decays exponentially due to the
exponential decrease of electric field produced by the oscillating charges in the GFM film.

We showed that magneto-electric coupling depends on the thickness of the FE layer.
For thin layers it grows linearly and saturates for thickness's exceeding some critical value.

\acknowledgments
I.~B. was supported by NSF under Cooperative Agreement Award EEC-1160504,
NSF Award DMR-1158666, and NSF PREM Award.

\bibliography{FE_mem}

\providecommand{\noopsort}[1]{}\providecommand{\singleletter}[1]{#1}%
\begin{thebibliography}{34}%
\makeatletter
\providecommand \@ifxundefined [1]{%
 \@ifx{#1\undefined}
}%
\providecommand \@ifnum [1]{%
 \ifnum #1\expandafter \@firstoftwo
 \else \expandafter \@secondoftwo
 \fi
}%
\providecommand \@ifx [1]{%
 \ifx #1\expandafter \@firstoftwo
 \else \expandafter \@secondoftwo
 \fi
}%
\providecommand \natexlab [1]{#1}%
\providecommand \enquote  [1]{``#1''}%
\providecommand \bibnamefont  [1]{#1}%
\providecommand \bibfnamefont [1]{#1}%
\providecommand \citenamefont [1]{#1}%
\providecommand \href@noop [0]{\@secondoftwo}%
\providecommand \href [0]{\begingroup \@sanitize@url \@href}%
\providecommand \@href[1]{\@@startlink{#1}\@@href}%
\providecommand \@@href[1]{\endgroup#1\@@endlink}%
\providecommand \@sanitize@url [0]{\catcode `\\12\catcode `\$12\catcode
  `\&12\catcode `\#12\catcode `\^12\catcode `\_12\catcode `\%12\relax}%
\providecommand \@@startlink[1]{}%
\providecommand \@@endlink[0]{}%
\providecommand \url  [0]{\begingroup\@sanitize@url \@url }%
\providecommand \@url [1]{\endgroup\@href {#1}{\urlprefix }}%
\providecommand \urlprefix  [0]{URL }%
\providecommand \Eprint [0]{\href }%
\providecommand \doibase [0]{http://dx.doi.org/}%
\providecommand \selectlanguage [0]{\@gobble}%
\providecommand \bibinfo  [0]{\@secondoftwo}%
\providecommand \bibfield  [0]{\@secondoftwo}%
\providecommand \translation [1]{[#1]}%
\providecommand \BibitemOpen [0]{}%
\providecommand \bibitemStop [0]{}%
\providecommand \bibitemNoStop [0]{.\EOS\space}%
\providecommand \EOS [0]{\spacefactor3000\relax}%
\providecommand \BibitemShut  [1]{\csname bibitem#1\endcsname}%
\let\auto@bib@innerbib\@empty
\bibitem [{\citenamefont {Eerenstein}\ \emph {et~al.}(2006)\citenamefont
  {Eerenstein}, \citenamefont {Mathur},\ and\ \citenamefont
  {Scott}}]{Scott2006}%
  \BibitemOpen
  \bibfield  {author} {\bibinfo {author} {\bibfnamefont {W.}~\bibnamefont
  {Eerenstein}}, \bibinfo {author} {\bibfnamefont {N.~D.}\ \bibnamefont
  {Mathur}}, \ and\ \bibinfo {author} {\bibfnamefont {J.~F.}\ \bibnamefont
  {Scott}},\ }\href@noop {} {\bibfield  {journal} {\bibinfo  {journal}
  {Nature}\ }\textbf {\bibinfo {volume} {442}},\ \bibinfo {pages} {759}
  (\bibinfo {year} {2006})}\BibitemShut {NoStop}%
\bibitem [{\citenamefont {Ramesh}\ and\ \citenamefont
  {Spaldin}(2007)}]{Spal2007}%
  \BibitemOpen
  \bibfield  {author} {\bibinfo {author} {\bibfnamefont {R.}~\bibnamefont
  {Ramesh}}\ and\ \bibinfo {author} {\bibfnamefont {N.~A.}\ \bibnamefont
  {Spaldin}},\ }\href@noop {} {\bibfield  {journal} {\bibinfo  {journal}
  {Nature Mat.}\ }\textbf {\bibinfo {volume} {6}},\ \bibinfo {pages} {21}
  (\bibinfo {year} {2007})}\BibitemShut {NoStop}%
\bibitem [{\citenamefont {Bibes}\ and\ \citenamefont
  {Barthelemy}(2008)}]{Bar2008}%
  \BibitemOpen
  \bibfield  {author} {\bibinfo {author} {\bibfnamefont {M.}~\bibnamefont
  {Bibes}}\ and\ \bibinfo {author} {\bibfnamefont {A.}~\bibnamefont
  {Barthelemy}},\ }\href@noop {} {\bibfield  {journal} {\bibinfo  {journal}
  {Nature Mat.}\ }\textbf {\bibinfo {volume} {7}},\ \bibinfo {pages} {425}
  (\bibinfo {year} {2008})}\BibitemShut {NoStop}%
\bibitem [{\citenamefont {Ohno}\ \emph {et~al.}(2000)\citenamefont {Ohno},
  \citenamefont {Chiba}, \citenamefont {Matsukura}, \citenamefont {Omiya},
  \citenamefont {Abe}, \citenamefont {Dietl}, \citenamefont {Ohno},\ and\
  \citenamefont {Ohtani}}]{Ohtani2000}%
  \BibitemOpen
  \bibfield  {author} {\bibinfo {author} {\bibfnamefont {H.}~\bibnamefont
  {Ohno}}, \bibinfo {author} {\bibfnamefont {D.}~\bibnamefont {Chiba}},
  \bibinfo {author} {\bibfnamefont {F.}~\bibnamefont {Matsukura}}, \bibinfo
  {author} {\bibfnamefont {T.}~\bibnamefont {Omiya}}, \bibinfo {author}
  {\bibfnamefont {E.}~\bibnamefont {Abe}}, \bibinfo {author} {\bibfnamefont
  {T.}~\bibnamefont {Dietl}}, \bibinfo {author} {\bibfnamefont
  {Y.}~\bibnamefont {Ohno}}, \ and\ \bibinfo {author} {\bibfnamefont
  {K.}~\bibnamefont {Ohtani}},\ }\href@noop {} {\bibfield  {journal} {\bibinfo
  {journal} {Nature (London)}\ }\textbf {\bibinfo {volume} {408}},\ \bibinfo
  {pages} {944} (\bibinfo {year} {2000})}\BibitemShut {NoStop}%
\bibitem [{\citenamefont {Chiba}\ \emph {et~al.}(2008)\citenamefont {Chiba},
  \citenamefont {Sawicki}, \citenamefont {Nishitani}, \citenamefont {Nakatani},
  \citenamefont {Matsukura},\ and\ \citenamefont {Ohno}}]{Ohno2008}%
  \BibitemOpen
  \bibfield  {author} {\bibinfo {author} {\bibfnamefont {D.}~\bibnamefont
  {Chiba}}, \bibinfo {author} {\bibfnamefont {M.}~\bibnamefont {Sawicki}},
  \bibinfo {author} {\bibfnamefont {Y.}~\bibnamefont {Nishitani}}, \bibinfo
  {author} {\bibfnamefont {Y.}~\bibnamefont {Nakatani}}, \bibinfo {author}
  {\bibfnamefont {F.}~\bibnamefont {Matsukura}}, \ and\ \bibinfo {author}
  {\bibfnamefont {H.}~\bibnamefont {Ohno}},\ }\href@noop {} {\bibfield
  {journal} {\bibinfo  {journal} {Nature (London)}\ }\textbf {\bibinfo {volume}
  {455}},\ \bibinfo {pages} {515} (\bibinfo {year} {2008})}\BibitemShut
  {NoStop}%
\bibitem [{\citenamefont {Chiba}\ \emph {et~al.}(2003)\citenamefont {Chiba},
  \citenamefont {Yamanouchi}, \citenamefont {Matsukura},\ and\ \citenamefont
  {Ohno}}]{Ohno2003}%
  \BibitemOpen
  \bibfield  {author} {\bibinfo {author} {\bibfnamefont {D.}~\bibnamefont
  {Chiba}}, \bibinfo {author} {\bibfnamefont {M.}~\bibnamefont {Yamanouchi}},
  \bibinfo {author} {\bibfnamefont {F.}~\bibnamefont {Matsukura}}, \ and\
  \bibinfo {author} {\bibfnamefont {H.}~\bibnamefont {Ohno}},\ }\href@noop {}
  {\bibfield  {journal} {\bibinfo  {journal} {Science}\ }\textbf {\bibinfo
  {volume} {301}},\ \bibinfo {pages} {943} (\bibinfo {year}
  {2003})}\BibitemShut {NoStop}%
\bibitem [{\citenamefont {Katsura}\ \emph {et~al.}(2005)\citenamefont
  {Katsura}, \citenamefont {Nagaosa},\ and\ \citenamefont
  {Balatsky}}]{Bal2005}%
  \BibitemOpen
  \bibfield  {author} {\bibinfo {author} {\bibfnamefont {H.}~\bibnamefont
  {Katsura}}, \bibinfo {author} {\bibfnamefont {N.}~\bibnamefont {Nagaosa}}, \
  and\ \bibinfo {author} {\bibfnamefont {A.~V.}\ \bibnamefont {Balatsky}},\
  }\href@noop {} {\bibfield  {journal} {\bibinfo  {journal} {Phys.\ Rev.\
  Lett}\ }\textbf {\bibinfo {volume} {95}},\ \bibinfo {pages} {057205}
  (\bibinfo {year} {2005})}\BibitemShut {NoStop}%
\bibitem [{\citenamefont {Sergienko}\ and\ \citenamefont
  {Dagotto}(2006)}]{Ser2006}%
  \BibitemOpen
  \bibfield  {author} {\bibinfo {author} {\bibfnamefont {I.~A.}\ \bibnamefont
  {Sergienko}}\ and\ \bibinfo {author} {\bibfnamefont {E.}~\bibnamefont
  {Dagotto}},\ }\href@noop {} {\bibfield  {journal} {\bibinfo  {journal}
  {Phys.\ Rev.\ B}\ }\textbf {\bibinfo {volume} {73}},\ \bibinfo {pages}
  {094434} (\bibinfo {year} {2006})}\BibitemShut {NoStop}%
\bibitem [{\citenamefont {Nan}(1994)}]{Nan1994}%
  \BibitemOpen
  \bibfield  {author} {\bibinfo {author} {\bibfnamefont {C.-W.}\ \bibnamefont
  {Nan}},\ }\href@noop {} {\bibfield  {journal} {\bibinfo  {journal} {Phys.\
  Rev.\ B}\ }\textbf {\bibinfo {volume} {50}},\ \bibinfo {pages} {6082}
  (\bibinfo {year} {1994})}\BibitemShut {NoStop}%
\bibitem [{\citenamefont {Thiele}\ \emph {et~al.}(2007)\citenamefont {Thiele},
  \citenamefont {Dorr}, \citenamefont {Bilani}, \citenamefont {Rodel},\ and\
  \citenamefont {Schultz}}]{Schultz2007}%
  \BibitemOpen
  \bibfield  {author} {\bibinfo {author} {\bibfnamefont {C.}~\bibnamefont
  {Thiele}}, \bibinfo {author} {\bibfnamefont {K.}~\bibnamefont {Dorr}},
  \bibinfo {author} {\bibfnamefont {O.}~\bibnamefont {Bilani}}, \bibinfo
  {author} {\bibfnamefont {J.}~\bibnamefont {Rodel}}, \ and\ \bibinfo {author}
  {\bibfnamefont {L.}~\bibnamefont {Schultz}},\ }\href@noop {} {\bibfield
  {journal} {\bibinfo  {journal} {Phys.\ Rev.\ B}\ }\textbf {\bibinfo {volume}
  {75}},\ \bibinfo {pages} {054408} (\bibinfo {year} {2007})}\BibitemShut
  {NoStop}%
\bibitem [{\citenamefont {Geprags}\ \emph {et~al.}(2010)\citenamefont
  {Geprags}, \citenamefont {Brandlmaier}, \citenamefont {Opel}, \citenamefont
  {Gross},\ and\ \citenamefont {Goennenwein}}]{Goennenwein2010}%
  \BibitemOpen
  \bibfield  {author} {\bibinfo {author} {\bibfnamefont {S.}~\bibnamefont
  {Geprags}}, \bibinfo {author} {\bibfnamefont {A.}~\bibnamefont
  {Brandlmaier}}, \bibinfo {author} {\bibfnamefont {M.}~\bibnamefont {Opel}},
  \bibinfo {author} {\bibfnamefont {R.}~\bibnamefont {Gross}}, \ and\ \bibinfo
  {author} {\bibfnamefont {S.~T.~B.}\ \bibnamefont {Goennenwein}},\ }\href@noop
  {} {\bibfield  {journal} {\bibinfo  {journal} {Appl. Phys. Lett.}\ }\textbf
  {\bibinfo {volume} {96}},\ \bibinfo {pages} {142509} (\bibinfo {year}
  {2010})}\BibitemShut {NoStop}%
\bibitem [{\citenamefont {Weisheit}\ \emph {et~al.}(2007)\citenamefont
  {Weisheit}, \citenamefont {Fahler}, \citenamefont {Marty}, \citenamefont
  {Souche}, \citenamefont {Poinsignon},\ and\ \citenamefont
  {Givord}}]{Givord2007}%
  \BibitemOpen
  \bibfield  {author} {\bibinfo {author} {\bibfnamefont {M.}~\bibnamefont
  {Weisheit}}, \bibinfo {author} {\bibfnamefont {S.}~\bibnamefont {Fahler}},
  \bibinfo {author} {\bibfnamefont {A.}~\bibnamefont {Marty}}, \bibinfo
  {author} {\bibfnamefont {Y.}~\bibnamefont {Souche}}, \bibinfo {author}
  {\bibfnamefont {C.}~\bibnamefont {Poinsignon}}, \ and\ \bibinfo {author}
  {\bibfnamefont {D.}~\bibnamefont {Givord}},\ }\href@noop {} {\bibfield
  {journal} {\bibinfo  {journal} {Science}\ }\textbf {\bibinfo {volume}
  {315}},\ \bibinfo {pages} {349} (\bibinfo {year} {2007})}\BibitemShut
  {NoStop}%
\bibitem [{\citenamefont {Tsujikawa}\ and\ \citenamefont
  {Oda}(2009)}]{Oda2009}%
  \BibitemOpen
  \bibfield  {author} {\bibinfo {author} {\bibfnamefont {M.}~\bibnamefont
  {Tsujikawa}}\ and\ \bibinfo {author} {\bibfnamefont {T.}~\bibnamefont
  {Oda}},\ }\href@noop {} {\bibfield  {journal} {\bibinfo  {journal} {Phys.
  Rev. Lett.}\ }\textbf {\bibinfo {volume} {102}},\ \bibinfo {pages} {247203}
  (\bibinfo {year} {2009})}\BibitemShut {NoStop}%
\bibitem [{\citenamefont {Duan}\ \emph {et~al.}(2008)\citenamefont {Duan},
  \citenamefont {Velev}, \citenamefont {Sabirianov}, \citenamefont {Zhu},
  \citenamefont {Chu}, \citenamefont {Jaswal},\ and\ \citenamefont
  {Tsymbal}}]{Tsymbal2008}%
  \BibitemOpen
  \bibfield  {author} {\bibinfo {author} {\bibfnamefont {C.-G.}\ \bibnamefont
  {Duan}}, \bibinfo {author} {\bibfnamefont {J.~P.}\ \bibnamefont {Velev}},
  \bibinfo {author} {\bibfnamefont {R.~F.}\ \bibnamefont {Sabirianov}},
  \bibinfo {author} {\bibfnamefont {Z.}~\bibnamefont {Zhu}}, \bibinfo {author}
  {\bibfnamefont {J.}~\bibnamefont {Chu}}, \bibinfo {author} {\bibfnamefont
  {S.~S.}\ \bibnamefont {Jaswal}}, \ and\ \bibinfo {author} {\bibfnamefont
  {E.~Y.}\ \bibnamefont {Tsymbal}},\ }\href@noop {} {\bibfield  {journal}
  {\bibinfo  {journal} {Phys. Rev. Lett.}\ }\textbf {\bibinfo {volume} {101}},\
  \bibinfo {pages} {137201} (\bibinfo {year} {2008})}\BibitemShut {NoStop}%
\bibitem [{\citenamefont {Zhuravlev}\ \emph {et~al.}(2010)\citenamefont
  {Zhuravlev}, \citenamefont {Maekawa},\ and\ \citenamefont
  {Tsymbal}}]{Tsymbal2010}%
  \BibitemOpen
  \bibfield  {author} {\bibinfo {author} {\bibfnamefont {M.~Y.}\ \bibnamefont
  {Zhuravlev}}, \bibinfo {author} {\bibfnamefont {S.}~\bibnamefont {Maekawa}},
  \ and\ \bibinfo {author} {\bibfnamefont {E.~Y.}\ \bibnamefont {Tsymbal}},\
  }\href@noop {} {\bibfield  {journal} {\bibinfo  {journal} {Phys.\ Rev.\ B}\
  }\textbf {\bibinfo {volume} {81}},\ \bibinfo {pages} {104419} (\bibinfo
  {year} {2010})}\BibitemShut {NoStop}%
\bibitem [{\citenamefont {Garcia}\ \emph {et~al.}(2010)\citenamefont {Garcia},
  \citenamefont {Bibes}, \citenamefont {Bocher}, \citenamefont {Valencia},
  \citenamefont {Kronast}, \citenamefont {Crassous}, \citenamefont {Moya},
  \citenamefont {Enouz-Vedrenne}, \citenamefont {Gloter}, \citenamefont
  {Imhoff}, \citenamefont {Deranlot}, \citenamefont {Mathur}, \citenamefont
  {Fusil}, \citenamefont {Bouzehouane},\ and\ \citenamefont
  {Barthelemy}}]{Barthelemy2010}%
  \BibitemOpen
  \bibfield  {author} {\bibinfo {author} {\bibfnamefont {V.}~\bibnamefont
  {Garcia}}, \bibinfo {author} {\bibfnamefont {M.}~\bibnamefont {Bibes}},
  \bibinfo {author} {\bibfnamefont {L.}~\bibnamefont {Bocher}}, \bibinfo
  {author} {\bibfnamefont {S.}~\bibnamefont {Valencia}}, \bibinfo {author}
  {\bibfnamefont {F.}~\bibnamefont {Kronast}}, \bibinfo {author} {\bibfnamefont
  {A.}~\bibnamefont {Crassous}}, \bibinfo {author} {\bibfnamefont
  {X.}~\bibnamefont {Moya}}, \bibinfo {author} {\bibfnamefont {S.}~\bibnamefont
  {Enouz-Vedrenne}}, \bibinfo {author} {\bibfnamefont {A.}~\bibnamefont
  {Gloter}}, \bibinfo {author} {\bibfnamefont {D.}~\bibnamefont {Imhoff}},
  \bibinfo {author} {\bibfnamefont {C.}~\bibnamefont {Deranlot}}, \bibinfo
  {author} {\bibfnamefont {N.~D.}\ \bibnamefont {Mathur}}, \bibinfo {author}
  {\bibfnamefont {S.}~\bibnamefont {Fusil}}, \bibinfo {author} {\bibfnamefont
  {K.}~\bibnamefont {Bouzehouane}}, \ and\ \bibinfo {author} {\bibfnamefont
  {A.}~\bibnamefont {Barthelemy}},\ }\href@noop {} {\bibfield  {journal}
  {\bibinfo  {journal} {Science}\ }\textbf {\bibinfo {volume} {327}},\ \bibinfo
  {pages} {1106} (\bibinfo {year} {2010})}\BibitemShut {NoStop}%
\bibitem [{\citenamefont {Jia}\ and\ \citenamefont
  {Berakdar}(2009)}]{Berakdar2009}%
  \BibitemOpen
  \bibfield  {author} {\bibinfo {author} {\bibfnamefont {C.}~\bibnamefont
  {Jia}}\ and\ \bibinfo {author} {\bibfnamefont {J.}~\bibnamefont {Berakdar}},\
  }\href@noop {} {\bibfield  {journal} {\bibinfo  {journal} {Phys.\ Rev.\ B}\
  }\textbf {\bibinfo {volume} {80}},\ \bibinfo {pages} {014432} (\bibinfo
  {year} {2009})}\BibitemShut {NoStop}%
\bibitem [{\citenamefont {Jia}\ and\ \citenamefont
  {Berakdar}(2011)}]{Berakdar2011}%
  \BibitemOpen
  \bibfield  {author} {\bibinfo {author} {\bibfnamefont {C.}~\bibnamefont
  {Jia}}\ and\ \bibinfo {author} {\bibfnamefont {J.}~\bibnamefont {Berakdar}},\
  }\href@noop {} {\bibfield  {journal} {\bibinfo  {journal} {Phys.\ Rev.\ B}\
  }\textbf {\bibinfo {volume} {83}},\ \bibinfo {pages} {045309} (\bibinfo
  {year} {2011})}\BibitemShut {NoStop}%
\bibitem [{\citenamefont {Udalov}\ \emph
  {et~al.}(2014{\natexlab{a}})\citenamefont {Udalov}, \citenamefont
  {Chtchelkatchev},\ and\ \citenamefont {Beloborodov}}]{Beloborodov2014_ME}%
  \BibitemOpen
  \bibfield  {author} {\bibinfo {author} {\bibfnamefont {O.~G.}\ \bibnamefont
  {Udalov}}, \bibinfo {author} {\bibfnamefont {N.~M.}\ \bibnamefont
  {Chtchelkatchev}}, \ and\ \bibinfo {author} {\bibfnamefont {I.~S.}\
  \bibnamefont {Beloborodov}},\ }\href@noop {} {\bibfield  {journal} {\bibinfo
  {journal} {Phys.\ Rev.\ B}\ }\textbf {\bibinfo {volume} {89}},\ \bibinfo
  {pages} {174203} (\bibinfo {year} {2014}{\natexlab{a}})}\BibitemShut
  {NoStop}%
\bibitem [{\citenamefont {Udalov}\ \emph
  {et~al.}(2014{\natexlab{b}})\citenamefont {Udalov}, \citenamefont
  {Chtchelkatchev},\ and\ \citenamefont {Beloborodov}}]{Beloborodov2014_ME1}%
  \BibitemOpen
  \bibfield  {author} {\bibinfo {author} {\bibfnamefont {O.~G.}\ \bibnamefont
  {Udalov}}, \bibinfo {author} {\bibfnamefont {N.~M.}\ \bibnamefont
  {Chtchelkatchev}}, \ and\ \bibinfo {author} {\bibfnamefont {I.~S.}\
  \bibnamefont {Beloborodov}},\ }\href@noop {} {\  (\bibinfo {year}
  {2014}{\natexlab{b}})},\ \Eprint {http://arxiv.org/abs/1404.6671}
  {arXiv:1404.6671 [cond-mat]} \BibitemShut {NoStop}%
\bibitem [{\citenamefont {Smolenskii}\ and\ \citenamefont
  {Chupis}(1982)}]{Chupis1982}%
  \BibitemOpen
  \bibfield  {author} {\bibinfo {author} {\bibfnamefont {G.~A.}\ \bibnamefont
  {Smolenskii}}\ and\ \bibinfo {author} {\bibfnamefont {I.~E.}\ \bibnamefont
  {Chupis}},\ }\href@noop {} {\bibfield  {journal} {\bibinfo  {journal} {Sov.
  Phys. Usp.}\ }\textbf {\bibinfo {volume} {25}},\ \bibinfo {pages} {475}
  (\bibinfo {year} {1982})}\BibitemShut {NoStop}%
\bibitem [{\citenamefont {Landau}\ and\ \citenamefont
  {Lifshitz}(1960)}]{landauVol8}%
  \BibitemOpen
  \bibfield  {author} {\bibinfo {author} {\bibfnamefont {L.~D.}\ \bibnamefont
  {Landau}}\ and\ \bibinfo {author} {\bibfnamefont {E.}~\bibnamefont
  {Lifshitz}},\ }\href@noop {} {\emph {\bibinfo {title} {Course of Theoretical
  Physics: Vol.: 8: Electrodynamics of Continuous Medie}}}\ (\bibinfo
  {publisher} {Pergamon Press},\ \bibinfo {year} {1960})\BibitemShut {NoStop}%
\bibitem [{\citenamefont {Devonshire}(1949)}]{devonshire1949xcvi}%
  \BibitemOpen
  \bibfield  {author} {\bibinfo {author} {\bibfnamefont {A.~F.}\ \bibnamefont
  {Devonshire}},\ }\href@noop {} {\bibfield  {journal} {\bibinfo  {journal}
  {Philosophical Magazine}\ }\textbf {\bibinfo {volume} {40}},\ \bibinfo
  {pages} {1040} (\bibinfo {year} {1949})}\BibitemShut {NoStop}%
\bibitem [{\citenamefont {Strukov}\ and\ \citenamefont
  {Levanyuk}(1998)}]{Levan1983}%
  \BibitemOpen
  \bibfield  {author} {\bibinfo {author} {\bibfnamefont {B.~A.}\ \bibnamefont
  {Strukov}}\ and\ \bibinfo {author} {\bibfnamefont {A.~P.}\ \bibnamefont
  {Levanyuk}},\ }\href@noop {} {\emph {\bibinfo {title} {Ferroelectric
  Phenomena in Crystals}}}\ (\bibinfo  {publisher} {Springer, Geidelberg,
  1998},\ \bibinfo {year} {1998})\BibitemShut {NoStop}%
\bibitem [{\citenamefont {Ong}\ \emph {et~al.}(2001)\citenamefont {Ong},
  \citenamefont {Osman},\ and\ \citenamefont {Tilley}}]{Tilley2001}%
  \BibitemOpen
  \bibfield  {author} {\bibinfo {author} {\bibfnamefont {L.-H.}\ \bibnamefont
  {Ong}}, \bibinfo {author} {\bibfnamefont {J.}~\bibnamefont {Osman}}, \ and\
  \bibinfo {author} {\bibfnamefont {D.~R.}\ \bibnamefont {Tilley}},\
  }\href@noop {} {\bibfield  {journal} {\bibinfo  {journal} {Phys. Rev. B}\
  }\textbf {\bibinfo {volume} {63}},\ \bibinfo {pages} {144109} (\bibinfo
  {year} {2001})}\BibitemShut {NoStop}%
\bibitem [{\citenamefont {Chandra}\ and\ \citenamefont
  {Littlewood}(2007)}]{chandra2007landau}%
  \BibitemOpen
  \bibfield  {author} {\bibinfo {author} {\bibfnamefont {P.}~\bibnamefont
  {Chandra}}\ and\ \bibinfo {author} {\bibfnamefont {P.~B.}\ \bibnamefont
  {Littlewood}},\ }in\ \href@noop {} {\emph {\bibinfo {booktitle} {Physics of
  Ferroelectrics}}}\ (\bibinfo  {publisher} {Springer},\ \bibinfo {year}
  {2007})\ pp.\ \bibinfo {pages} {69--116}\BibitemShut {NoStop}%
\bibitem [{\citenamefont {Gittleman}\ \emph {et~al.}(1972)\citenamefont
  {Gittleman}, \citenamefont {Goldstein},\ and\ \citenamefont
  {Bozowski}}]{Boz1972}%
  \BibitemOpen
  \bibfield  {author} {\bibinfo {author} {\bibfnamefont {J.~I.}\ \bibnamefont
  {Gittleman}}, \bibinfo {author} {\bibfnamefont {Y.}~\bibnamefont
  {Goldstein}}, \ and\ \bibinfo {author} {\bibfnamefont {S.}~\bibnamefont
  {Bozowski}},\ }\href@noop {} {\bibfield  {journal} {\bibinfo  {journal}
  {Phys. Rev. B}\ }\textbf {\bibinfo {volume} {5}},\ \bibinfo {pages} {3609}
  (\bibinfo {year} {1972})}\BibitemShut {NoStop}%
\bibitem [{\citenamefont {Barzilai}\ \emph {et~al.}(1981)\citenamefont
  {Barzilai}, \citenamefont {Goldstein}, \citenamefont {Balberg},\ and\
  \citenamefont {Helman}}]{Hel1981}%
  \BibitemOpen
  \bibfield  {author} {\bibinfo {author} {\bibfnamefont {S.}~\bibnamefont
  {Barzilai}}, \bibinfo {author} {\bibfnamefont {Y.}~\bibnamefont {Goldstein}},
  \bibinfo {author} {\bibfnamefont {I.}~\bibnamefont {Balberg}}, \ and\
  \bibinfo {author} {\bibfnamefont {J.~S.}\ \bibnamefont {Helman}},\
  }\href@noop {} {\bibfield  {journal} {\bibinfo  {journal} {Phys. Rev. B}\
  }\textbf {\bibinfo {volume} {23}},\ \bibinfo {pages} {1809} (\bibinfo {year}
  {1981})}\BibitemShut {NoStop}%
\bibitem [{\citenamefont {Kimura}\ and\ \citenamefont
  {Ohigashi}(1983)}]{Ohigashi1983}%
  \BibitemOpen
  \bibfield  {author} {\bibinfo {author} {\bibfnamefont {K.}~\bibnamefont
  {Kimura}}\ and\ \bibinfo {author} {\bibfnamefont {H.}~\bibnamefont
  {Ohigashi}},\ }\href@noop {} {\bibfield  {journal} {\bibinfo  {journal}
  {Appl. Phys. Lett.}\ }\textbf {\bibinfo {volume} {43}},\ \bibinfo {pages}
  {834} (\bibinfo {year} {1983})}\BibitemShut {NoStop}%
\bibitem [{\citenamefont {Yamada}\ and\ \citenamefont
  {Kitayama}(1981)}]{Kitayama1981}%
  \BibitemOpen
  \bibfield  {author} {\bibinfo {author} {\bibfnamefont {T.}~\bibnamefont
  {Yamada}}\ and\ \bibinfo {author} {\bibfnamefont {T.}~\bibnamefont
  {Kitayama}},\ }\href@noop {} {\bibfield  {journal} {\bibinfo  {journal} {J.
  Appl. Phys.}\ }\textbf {\bibinfo {volume} {52}},\ \bibinfo {pages} {6859}
  (\bibinfo {year} {1981})}\BibitemShut {NoStop}%
\bibitem [{\citenamefont {Torrance}\ \emph {et~al.}(1981)\citenamefont
  {Torrance}, \citenamefont {Vazquez}, \citenamefont {Mayerle}, ,\ and\
  \citenamefont {Lee}}]{Lee1981}%
  \BibitemOpen
  \bibfield  {author} {\bibinfo {author} {\bibfnamefont {J.~B.}\ \bibnamefont
  {Torrance}}, \bibinfo {author} {\bibfnamefont {J.~E.}\ \bibnamefont
  {Vazquez}}, \bibinfo {author} {\bibfnamefont {J.~J.}\ \bibnamefont
  {Mayerle}}, , \ and\ \bibinfo {author} {\bibfnamefont {V.~Y.}\ \bibnamefont
  {Lee}},\ }\href@noop {} {\bibfield  {journal} {\bibinfo  {journal} {Phys.
  Rev. Lett.}\ }\textbf {\bibinfo {volume} {46}},\ \bibinfo {pages} {253}
  (\bibinfo {year} {1981})}\BibitemShut {NoStop}%
\bibitem [{\citenamefont {Berg}\ and\ \citenamefont
  {Avoird}(1989)}]{Avoird1981}%
  \BibitemOpen
  \bibfield  {author} {\bibinfo {author} {\bibfnamefont {T.~H. M. V.~D.}\
  \bibnamefont {Berg}}\ and\ \bibinfo {author} {\bibfnamefont {A.~V.~D.}\
  \bibnamefont {Avoird}},\ }\href@noop {} {\bibfield  {journal} {\bibinfo
  {journal} {Chem. Phys. Lett.}\ }\textbf {\bibinfo {volume} {160}},\ \bibinfo
  {pages} {223} (\bibinfo {year} {1989})}\BibitemShut {NoStop}%
\bibitem [{\citenamefont {Fridkin}(2006)}]{Frid2006rev}%
  \BibitemOpen
  \bibfield  {author} {\bibinfo {author} {\bibfnamefont {V.~M.}\ \bibnamefont
  {Fridkin}},\ }\href@noop {} {\bibfield  {journal} {\bibinfo  {journal} {Sov.
  Phys. Usp.}\ }\textbf {\bibinfo {volume} {49}},\ \bibinfo {pages} {193}
  (\bibinfo {year} {2006})}\BibitemShut {NoStop}%
\bibitem [{\citenamefont {Fridkin}\ \emph {et~al.}(2010)\citenamefont
  {Fridkin}, \citenamefont {Gaynutdinov},\ and\ \citenamefont
  {Ducharme}}]{Frid2010rev}%
  \BibitemOpen
  \bibfield  {author} {\bibinfo {author} {\bibfnamefont {V.~M.}\ \bibnamefont
  {Fridkin}}, \bibinfo {author} {\bibfnamefont {R.~V.}\ \bibnamefont
  {Gaynutdinov}}, \ and\ \bibinfo {author} {\bibfnamefont {S.}~\bibnamefont
  {Ducharme}},\ }\href@noop {} {\bibfield  {journal} {\bibinfo  {journal} {Sov.
  Phys. Usp.}\ }\textbf {\bibinfo {volume} {53}},\ \bibinfo {pages} {199}
  (\bibinfo {year} {2010})}\BibitemShut {NoStop}%
\end{thebibliography}%

\end{document}